\documentclass[%
 reprint,
 amsmath,amssymb,
 aps,
]{revtex4-2}

\usepackage{graphicx}
\usepackage{dcolumn}
\usepackage{bm}
\usepackage{dutchcal}
\usepackage{comment}
\usepackage{tabularx}
\usepackage{dcolumn}
\usepackage{bm}
\usepackage[utf8]{inputenc}
\usepackage{amsfonts}
\usepackage{caption}
\usepackage{xcolor}
\usepackage{subcaption}
\usepackage{graphicx}
\usepackage{dcolumn}
\usepackage{bm}

\begin{document}

\preprint{APS/123-QED}

\title{A rigorous study on the longitudinal beam coupling impedance of a cylindrical lossy pipe in normal and anomalous regimes}

\author{A. Curcio, M. Migliorati, A. Mostacci, L. Palumbo}




\date{\today}

\begin{abstract}
One of the primary issues in designing particle accelerators is the effect of energy losses and collective instabilities caused by the conductivity of the vacuum chamber. Due to its relevance, many authors have long focused on studying the coupling impedance of lossy pipes with various geometries and metals. Most studies were developed for relativistic beams and room-temperature conductivity. In recent years, there has been increasing interest in the high-frequency impedance behavior of ultra-short bunches, as in the case of FELs,  and in the anomalous conductivity at cryogenic temperatures of the vacuum chambers.
This work introduces a new unifying theoretical framework for analyzing electromagnetic interactions in cylindrical pipes excited by charged particles traveling on the axis. In this context, a key achievement is the derivation of a rigorous expression for the electric conductivity, based on the Boltzmann theory of a Fermi gas, which forms the foundation for subsequent developments. New formulas for the surface impedance of cylindrical pipes are presented, from which a more general impedance expression is derived. Our approach captures the system's behavior across various energy and frequency regimes. 
Eventually, the anomalous regime is studied rigorously, providing analytical solutions for the electromagnetic field at cryogenic temperatures. Indeed, new and exact expressions for the anomalous surface impedance for carriers' reflection coefficients p = 0 and p = 1 are obtained, offering deeper insight into the electromagnetic properties of such systems under extreme conditions.\\
\end{abstract}

\maketitle


\section{Introduction}
The beam impedance \cite{vaccaro1966longitudinal,sessler1967longitudinal,Bane_Wilson_Weiland,Palumbo_Vaccaro,Weiland2,Chao_1993, Palumbo_Vaccaro_Zobov,Zotter_Keifets} of vacuum pipes with finite conductivity $\sigma$ (usually copper, steel, aluminum), also called resistive wall impedance, has been treated by many authors for various geometries (rectangular, elliptic, circular) and charge velocity $v=\beta c$ (with $c$ the speed of light in vacuum). Analytical models were developed for the DC (Direct Current) conductivity and AC (Alternate Current) conductivity cases \cite{Piwinski_flat,Palumbo_Vaccaro_lossy,Palumbo_Vaccaro,yokoya1993resistive,gluckstern1993coupling,Piwinski_elliptic,Ruggiero_1995_Arbitrary,bane1996short,Zimmermann_Oide,migliorati_FCC,stupakov2020resistive}, and for the so-called "anomalous regime" \cite{Stupakov}.
Semi-analytical models have also been developed for multilayer walls~\cite{wang2007resistive,Mounet_thesis, migliorati}.\\
Unlike in the ideal perfectly conductive case, the electromagnetic fields generated by a point charge traveling in a vacuum chamber penetrate the wall, and the tangent electric field at the wall is not zero. The induced current dissipates energy in the material because of Joule's effect, which corresponds to an energy loss of the source charge. Compared to the perfectly conducting case, the pattern of the fields substantially differs. They remain behind the charge over a distance depending on the value of the electrical conductivity and the energy of the charge. \\
The present study is motivated by several fundamental questions concerning the validity and applicability of some classical electomagnetic models used in the literature. Specifically, a central issue is how finite particle energy affects both surface and coupling impedance, as well as energy losses, when the true cylindrical geometry of an accelerator pipe is taken into account. Moreover, we investigate whether and under what conditions the concept of anomalous surface impedance, originally developed by Reuter and Sondheimer \cite{reuter1948theory} for electromagnetic waves impinging on flat surfaces, can be meaningfully extended to the case of a cylindrical pipe excited by a Coulombian field of a moving point charge.
\\

In our study, we assume the wall conductivity is so high that the depth of the skin is much smaller than the thickness of the pipe. In deriving fields, we apply the "Green function" method introduced for a cylindrical lossy pipe in Refs. \cite{Palumbo_Vaccaro,Palumbo_Vaccaro_Zobov}, and applied in Refs. \cite{Zimmermann_Oide, stupakov2020resistive, al2001analytical, gluckstern2000analytic}. 
As a first step, we applied the Boltzmann transport equation \cite{grosso2013solid} to derive a more general expression of the electric conductivity in cylindrical pipe walls, which depends on the relaxation time, the geometry and the velocity of the charge generating the primary fields. We then obtained the surface impedance at the cylindrical pipe wall under normal and anomalous regimes. Further, we derived general expressions of the longitudinal beam impedance, showing the dependence on the energy for copper, aluminum and steel.
Finally, we discussed the results obtained for the wake potential in the case of copper, varying the energy in both normal and anomalous conductivity regimes.  
Whenever possible, the results are compared with the existing literature.
\\
To facilitate comparison with articles on the same topic, we recall that in the Refs. \cite{Chao_1993} and \cite{bane1996short} developed a theory of impedance and wakefields for infinite particle energy and normal surface impedance.
\\
A theory of coupling impedance for normal surface impedance at finite particle energy was developed by Refs. \cite{Palumbo_Vaccaro, stupakov2020resistive, al2001analytical, gluckstern2000analytic}.\\
In Refs. \cite{reuter1948theory} and \cite{dingle1953anomalous}, the authors developed models of anomalous surface impedance for a flat surface irradiated by an electromagnetic plane wave.\\
In Ref. \cite{Stupakov}, the authors use the anomalous surface impedance found in Refs. \cite{reuter1948theory} and \cite{dingle1953anomalous} to evaluate the coupling impedance in a cryogenic beam pipe for infinite particle energy.\\
In Ref. \cite{podobedov2009resistive}, the authors introduce a study of the impedance in the extremely anomalous regime of conductivity at large particle energies.
\\
In the present paper, we encompass all the above studies in a unique, rigorous frame valid for any particle energy and any regime of conductivity/surface impedance in a cylindrical pipe.

\section{Coordinate system and relevant definitions}

\subsection{Energy loss and wake function}
Let's consider a charge $q_0$ traveling with constant velocity $\vec{v}=\beta c \hat{z}$ on the axis $\hat{z}$ of a vacuum chamber with a circular cross-section. The charge $q_0$ at the position $s_0=\beta ct$, called $leading \ charge$,  generates electromagnetic fields during the flight in the vacuum chamber. Let us now consider a "test charge" $q$ moving on the axis with the same velocity but at the position $s=s_0+z$, where $z$ is the distance of the two charges. The parameter $z$ gives the distance of the test charge from the leading one, such that $z<0$ for a $trailing \  charge$. 
The energy variation of the test charge, obtained as the work done by the longitudinal electric force $F_z$ generated by the leading charge due to the interaction with the pipe walls,
\begin{equation}
\label{eq:Work_long_force}
    \mathcal{E}(z)=\int_{-\infty}^\infty  F_z(s_0, s=s_0+z) ds
  \end{equation}
is used to define the longitudinal wake potential \cite{Palumbo_Vaccaro_Zobov,Zotter_Keifets}:
\begin{equation}
\label{eq:wake_potential_definition}
    w_z(z)=-\frac{1}{q_0q}\mathcal{E}(z)=-\frac{1}{q_0q}\int_{-\infty}^\infty  F_z(s_0, s=s_0+z) ds
  \end{equation}
 which is positive for energy loss. For longitudinally uniform pipes, where $F_z$ depends only upon $z$, we introduce the energy variation per unit length for longitudinally uniform pipes:

\begin{equation}
\label{eq:Work_unit_length}
   \frac{d\mathcal{E}(z)}{ds}=   F_z(z) 
  \end{equation}
Accordingly, the wake function per unit length is defined as:
\begin{equation}
\label{eq:wake_per_umit_length}
    w'_z(z)=\frac{d w_z(z)}{ds}=-\frac{1}{q_0q}\frac{d\mathcal{E}(z)}{ds}= -\frac{E_z(z)}{q_0}
\end{equation}

\subsection{Loss factor and beam loading theorem}
The leading charge $q_0$ can also be subject to the effects of its own fields that change its energy. The  $loss\ factor$   per unit length is the energy variation of the leading charge per unit length and unit charge and can be written as:
  \begin{equation}
\label{eq:loss_factor}
   k'=w'_z(0) =-\frac{E_z(z=0)}{q_0}
  \end{equation}
  The above definitions imply that the wake function gives the loss factor in the $z \rightarrow 0$ limit. This is true as long as $\beta<1$, however, in the relevant limit $\beta \rightarrow 1$,  it results in:

\begin{equation}
 \label{eq:beam_loading}
   w'_z(0)=\frac{1}{2}\lim_{z\to 0^+} \frac{d w_z(z)}{ds}=\frac{1}{2}w'_z(0^+)
\end{equation}
which expresses the beam loading theorem.
Indeed, when $\beta\rightarrow 1$, due to the causality, the fields vanish ahead of the leading charge, and the wake potential shows a discontinuity at $z=0$. 
\subsection{Wake potential of a charge distribution}

 A bunch is represented here by a distribution of charges moving as an ensemble with velocity $\vec{v}=\beta c \hat{z}$.  We describe the bunch as a linear charge distribution $\lambda(z)$, such that:

\begin{equation}
 Q_b=\int_{-\infty}^\infty \lambda(z)dz  
\end{equation}

The wake potential of a point charge is a Green function and allows us to compute the total wake potential generated by any distribution $\lambda(z)$. We split the bunch into infinitesimal slices of charge $dQ_b(z')=\lambda(z')dz'$, the energy change of this slice with a  charge $q$ at the position $z$ is: 
\begin{equation}
    d\mathcal{E}(z-z')=- q\lambda(z')w'_z(z-z')dz'ds
\end{equation}
By summing up the effects of all slices, normalizing to $Q_b$ we obtain the wake function $W'_z$ of the bunch distribution:
\begin{equation}
\label{eq:bunch_wake}
    W'_z(z)=-\frac{1}{qQ_b}\frac{d\mathcal{E}(z)}{ ds}=\frac{1}{Q_b}\int_{-\infty}^\infty  \lambda(z')w'_z(z-z')dz'
\end{equation}
In the ultra-relativistic limit $\beta\to 1$, the folding integral has $z$ as a lowermost limit due to the causality principle.

\subsection{Longitudinal beam impedance and loss factor}
The longitudinal impedance per unit length $Z'(\omega)$ is obtained from the frequency spectrum of the wake function:

\begin{equation}
\label{eq:longitudinal_impedance_definition_z}
Z'(\omega)=\frac{1}{\beta c}\int_{-\infty}^\infty w'_z(z) e^{i\frac{\omega z}{\beta c} }dz
\end{equation}
Namely, 
\begin{equation}
\label{eq:longitudinal_impedance_definition_z_E_z}
Z'(\omega)=-\frac{1}{q_0\beta c}\int_{-\infty}^\infty {E_z(z)}e^{i\frac{\omega z}{\beta c} }dz=-\frac{1}{q_0\beta c}\tilde{E}_z
(\omega)
\end{equation}
and
\begin{equation}
\label{wake_from_impedance}
    w'_z(z)=\frac{1}{2\pi }\int_{-\infty}^\infty Z'_r(\omega) e^{-i\frac{\omega z}{\beta c}}d\omega
\end{equation}
Eq. \ref{eq:longitudinal_impedance_definition_z_E_z} represents our definition of the Fourier transform, thus also establishing the relationship between the domain $z$ and the domain $\omega$, which in our work are reciprocals of each other because the phase velocity of the wakefield is equal to the speed of the leading particle.
The impedance is a complex quantity:
\begin{equation}
 Z'(\omega)=Z'_r(\omega)+i Z'_i(\omega) 
\end{equation} 
Since $w_z(z)$ is a real function, $Z_r(\omega)$ and $Z_i(\omega)$ are even and odd functions of $\omega$ respectively. From the above properties, it results that:

\begin{equation}
\label{k'}
    k'=\frac{1}{\pi}\int_0^\infty Z'_r(\omega)d\omega
\end{equation}
\subsection{Loss  factor and wake potential of a charge distribution}
The total energy lost by a bunch with a frequency spectrum $\tilde{\lambda}(\omega)$ is given by:

\begin{equation}
\label{K'_b}
    k'_{bunch}=\frac{1}{\pi Q_b^2}\int_0^\infty Z'_r(\omega)|\tilde{\lambda}(\omega)|^2d\omega
\end{equation}
where:

\begin{equation}
 \tilde{\lambda}(\omega)=\frac{1}{\beta c}\int_{-\infty}^\infty \lambda(z) e^{i\frac{\omega z}{\beta c} }dz   
\end{equation}
Finally, the wake potential  is given by:

\begin{equation}
\label{wakebunformula}
    W'_z(z)=\frac{1}{2\pi  Q_b}\int_{-\infty
    }^\infty Z'_r(\omega)\tilde{\lambda}(\omega)e^{-i\frac{\omega z}{\beta c}}d\omega
\end{equation}
\subsection{Surface impedance}

In this paper, we make use of the Leontovich  boundary condition \cite{Leontovich} to define the surface impedance $Z_s$ at the wall of the cylindrical beam pipe:
\begin{equation}
\label{eq:Z_s}
Z_{s}(\omega)=\frac{\tilde{E}_z(\omega)}{\tilde{H}_\varphi(\omega)}
\end{equation}
where $\tilde{E}_z(\omega)$ and $\tilde{H}_\varphi(\omega)$ are the time Fourier transform of the electric and magnetic field components, respectively, evaluated at the pipe surface.\\
The beam impedance and the loss factor will be expressed in terms of the surface impedance.

\section{ electrical conductivity inside the walls of the  cylindrical pipe  }
The electromagnetic fields in the pipe walls can be determined from the electrical conductivity properties of the medium.
In Appendix \ref{condusec}, we derive the general expression of the electrical conductivity for the case of a cylindrically symmetric conducting pipe excited by a point charge traveling at constant speed on its axis:
\begin{equation}
\label{gencond}
\sigma(k_r,\omega)= \frac{3 n_e e^2 \tau}{4 m  } \int_0^\pi  \frac{  \sin^3{\theta}d\theta }{1+i  \omega \tau-i k_r v_F \tau\cos{\theta} } 
\end{equation}
 where $n_e$ is the number of carriers per unit volume, $v_F$ is the Fermi velocity of the carriers, $\tau$ is the relaxation time (i.e. the mean time between two consecutive collisions of the carriers with the lattice ions), $m$ is the effective mass of the carriers in the conduction band, $e$ is the elementary charge.
Remarkably, the conductivity depends on the cylindrical symmetry and on the speed of the charge. In particular, the electric conductivity depends on the radial wave-vector $k_r$ of the fields propagating into the walls, which considers the carriers' non-local response in the conductive medium, associated with so-called anomalous effects \cite{grosso2013solid}.
We distinguish two regions of interest for the conductivity, which are relevant for calculating the fields inside the walls. These two regions are customarily called "normal" and "anomalous", depending on the value of the $|\mathcal{s}|$ parameter: 
\begin{equation}
    \mathcal{s}=\frac{-i k_r v_F \tau}{1+i \omega \tau},
\end{equation}
The normal region is obtained for $|\mathcal{s}|<<1$, where the electric conductivity can be expressed as:
\begin{equation}
\label{con0}
\sigma(\omega)\simeq \frac{3 n_e e^2 \tau}{4 m  } \int_0^\pi  \frac{  \sin^3{\theta}d\theta }{1+i  \omega \tau} 
\end{equation}
The integral in Eq. \ref{con0} can be analytically solved
and it tends to the classical conductivity of the Drude model:
\begin{equation}
\label{con1}
\sigma(\omega)\simeq \frac{ n_e e^2 \tau}{ m (1+i  \omega \tau) }=\varepsilon_0 \omega_p^2\left(\frac{  \tau}{  1+i  \omega \tau }\right)
\end{equation}
also expressed through the plasma frequency $\omega_p$.
For $|\mathcal{s}|\gtrsim1$, 
no approximations can be used; we are in the anomalous conductivity region where the skin depth is smaller than the electron mean path $v_F\tau$, and the general expression in Eq. \ref{gencond} must be used. \\
For $|\mathcal{s}|>>1$, the extremely anomalous region is accessed and  analytical treatments are possible \cite{podobedov2009resistive}.\\
Generally speaking, the response of the cylindrical pipe to the Coulomb field of the point charge travelling on-axis occurs, in general, through a conductivity which differs from the Drude model.
\\
In the normal regime, another relevant distinction can be made between DC (Direct Current) and AC (Alternating Current) conductivity. The DC approximation of the conductivity may be enough to perform a study when $\omega\tau<<1$ in Eq. \ref{con1}, thus obtaining:
\begin{equation}
\label{sigma0}
    \sigma_0\simeq\frac{ n_e e^2 \tau}{ m  }
\end{equation}
In the above case, the conductivity is purely real, accounting for the loss due to the collisions of the electrons with atoms.\\
If, on the contrary, $\omega\tau >>1$ the AC conductivity is approximated by
\begin{equation}
\label{sigmaAC}
    \sigma(\omega)\simeq-i\frac{ n_e e^2 }{m \omega  }
\end{equation}
which does not depend on the relaxation time. Indeed, when the frequency $\omega$ of the fields is very high, their oscillation period becomes very short compared to the relaxation time $\tau$. As a result, electrons cannot follow the fast variations of the electric field, they do not frequently collide with the lattice, and there are no losses.
The conductivity shows only an imaginary part that behaves like an inductance, which can store and exchange energy.

\section{Electromagnetic field in the vacuum pipe}
\label{sec1}
The charge density associated with a point charge moving along the main axis $z$  of a cylindrical beam pipe of radius $b$ at constant velocity $\beta c$ can be expressed, in cylindrical coordinates, as:
\begin{equation}
\label{moneq}
    \rho=q_0\frac{\delta(r)}{2\pi r}\delta(z-\beta c t)
\end{equation}
The equation for the scalar potential $\phi$ generated by the above monopolar term is:
\begin{equation}
\label{phieq0}
   \frac{\partial^2 \phi}{\partial r^2}+\frac{1}{r}\frac{\partial \phi}{\partial r}+\frac{1}{\gamma^2}\frac{\partial^2\phi}{\partial z^2}=-\frac{\rho}{\varepsilon_0}
\end{equation}
where $\varepsilon_0$ is the dielectric constant of vacuum and $\gamma=1/\sqrt{1-\beta^2}$ is the Lorentz factor of the particle, and we have assumed that $\phi=\phi(r,z-\beta c t)$, i.e. the phase velocity of the electromagnetic field carried by the point charge must be equal to its velocity.
The solution of Eq. \ref{phieq0} can be found through the following expansion of the scalar potential:
\begin{equation}
    \phi=\frac{1}{2\pi}\int \tilde{\phi}(r,\omega=k_z \beta c)e^{-i k_z(z-\beta c t)}dk_z
\end{equation}
where the angular frequency of the field is recognized to be $\omega=k_z \beta c$. 
 In the $(r,\omega)$ domain, the solution   can be written as \cite{Palumbo_Vaccaro}:
\begin{equation}
\label{phi00000}
\begin{split}
    \tilde{\phi}(r,\omega)&=\frac{q_0}{2\pi \varepsilon_0}\Bigg[K_0\left(\frac{\omega r}{\gamma\beta c}\right)-\frac{K_0\left(\frac{\omega b}{\gamma\beta c}\right)}{I_0\left(\frac{\omega b}{\gamma\beta c}\right)}I_0\left(\frac{\omega r}{\gamma\beta c}\right)+\\
    &+B_0^\sigma(\omega) I_0\left(\frac{\omega r}{\gamma\beta c}\right)\Bigg]
\end{split}    
\end{equation}
where $I_n(x)$ and $K_n(x)$  are the nth-order modified Bessel functions of the first and second kind, respectively.
 The first two terms represent the Coulomb field propagating in a perfectly conductive cylindrical pipe while, in the term proportional to $B^\sigma_0(\omega)$, we have isolated the resistivity effect, which must tend to zero for a perfectly conductive pipe.
The axial component $A_z$ of the vector potential (the only non-zero component for the monopolar term) can be found through the Lorentz gauge:
\begin{equation}
    \frac{\partial A_z}{\partial z}+\frac{1}{c^2}\frac{\partial \phi}{\partial t}=0
\end{equation}
Therefore, in the $(r,\omega)$ domain:
\begin{equation}
    \tilde{A}_z(r,\omega)= \frac{ \beta }{c}\tilde{\phi}(r,\omega)
\end{equation}
The electric field for $r<b$ is then calculated as:
\begin{equation}
\label{efi}
\tilde{E}_z(r,\omega)=-i k_z \beta c \tilde{A}_z(r,\omega)+i k_z \tilde{\phi}(r,\omega)=\frac{i \omega}{\gamma^2\beta c}\tilde{\phi}(r,\omega)
\end{equation}
For the cylindrical symmetry, we  also notice that:
\begin{equation}
\label{HE}
\tilde{H}_\varphi=-\frac{1}{\mu_0}\frac{\partial  \tilde{A}_z(r,\omega)}{\partial r}= -  \frac{ \beta }{\mu_0 c}\frac{\partial  \tilde{\phi}(r,\omega)}{\partial r}
\end{equation}
where $\mu_0=1/\varepsilon_0 c^2$ is the magnetic permeability of vacuum.
Therefore:
\begin{equation}
\begin{split}
\tilde{H}_\varphi(r,\omega)&= \frac{q_0 \omega}{2\pi \gamma}\Bigg[K_1\left(\frac{\omega r}{\gamma \beta c}\right)+\frac{K_0\left(\frac{\omega b}{\gamma\beta c}\right)}{I_0\left(\frac{\omega b}{\gamma\beta c}\right)}I_1\left(\frac{\omega r}{\gamma\beta c}\right)+\\
&-B^\sigma_0(\omega)I_1\left(\frac{\omega r}{\gamma\beta c}\right)\Bigg]
\end{split}
\end{equation}
We know that the Coulomb field inside a perfectly conductive pipe vanishes as $1/\gamma^2$ and, because of the absence of losses, the beam impedance is purely imaginary. Thus, our main interest regards the term with $B^\sigma_0(\omega)$, which gives an electric field on the axis  
\begin{equation}
    \tilde{E}^\sigma_z (0,\omega)=\frac{i q_0 \omega }{2\pi \varepsilon_0 \gamma^2\beta c}B^\sigma_0(\omega)
\end{equation}
and a  longitudinal beam impedance from (Eq. \ref{eq:longitudinal_impedance_definition_z_E_z}) 
\begin{equation}
  \label{eq:impedance_sigma} Z'^\sigma(\omega)=\frac{dZ^\sigma(\omega)}{ds}=-\frac{i  \omega Z_0}{2\pi c(\gamma\beta )^2}B^\sigma_0(\omega)
   \end{equation}
where $Z_0=(\varepsilon_0 c)^{-1}$ is the vacuum impedance.
\\
Here we adopted the same approach as in Ref. \cite{stupakov2020resistive} and previously in Ref. \cite{Palumbo_Vaccaro}, where the space charge part of the impedance is defined as the
 impedance in the same pipe in the limit of perfect conductivity. We have introduced a $B_0^\sigma$ parameter, which is zero for infinite conductivity. It is clear from our equations (see for example Eqs. \ref{phi00000} and \ref{efi} for the longitudinal electric field) that in the limit of infinite conductivity the only impedance that remains in the pipe is the space charge part. However, in the definition of longitudinal beam impedance (Eq. \ref{eq:impedance_sigma}) it is possible to notice that the latter is proportional to $B_0^\sigma$, therefore it does not include the space charge part.
\\
In the following, we derive the quantity $B^\sigma_0(\omega)$ for all the cases of interest in the normal and anomalous regimes. To this end, we use the general expression of electric conductivity, Eq. \ref{gencond} and its approximations, deriving fields inside the pipe's walls.

\section{Electromagnetic fields inside the pipe's walls: normal regime }
To study the fields in the normal conducting regime, we start with the wave equation for the electric
potential \cite{Palumbo_Vaccaro}:
\begin{equation}
\label{phieqnorm}
   \frac{\partial^2 \tilde{\phi}}{\partial r^2}+\frac{1}{r}\frac{\partial \tilde{\phi}}{\partial r}-\left(\frac{k_z^2}{\gamma^2}+i\mu_0 k_z\beta c \sigma(\omega)\right)\tilde{\phi}=0
\end{equation}
A general solution of Eq. \ref{phieqnorm} is found to be:
\begin{equation}
\label{phinorm000}
\tilde{\phi}=\frac{q_0 C_0^{\sigma}(\omega)}{2\pi \varepsilon_0}K_0\left(\bar{\eta}r\right)   
\end{equation}
in which we have introduced the unknown quantity $C_0^{\sigma}(\omega)$ that guarantees the continuity of the electric field at the vacuum-pipe interface, and we defined:
\begin{equation}
    \bar{\eta}=\frac{\omega}{\beta c}\sqrt{\frac{1}{ \gamma^2}+i\frac{\beta^2\sigma(\omega)}{\varepsilon_0\omega} }
\end{equation}
The Lorentz gauge in the resistive medium of the pipe walls is expressed  as:
\begin{equation}
\label{lorgaug2}
\tilde{A}_z(r,\omega)= \frac{\left(1-\frac{i\sigma(\omega)}{\varepsilon_0 \omega}\right) \beta }{c}\tilde{\phi}(r,\omega)
\end{equation}
Furthermore, the electric field inside the pipe walls can be calculated using Eq. \ref{lorgaug2}, and it is:
\begin{equation}
\label{efi2}
\begin{split}
\tilde{E}_z(r,\omega)
 = \frac{\omega }{ c}
\left(\frac{i}{\beta^2\gamma^2}-\frac{\sigma(\omega)}{\varepsilon_0 \omega}\right)\frac{q_0 \beta}{2\pi \varepsilon_0}C_0^\sigma(\omega)K_0\left(\bar{\eta}r\right)
\end{split}
\end{equation}
Therefore, the magnetic field is found to be:
\begin{equation}
\label{HE2}
\begin{split}
\tilde{H}_\varphi(r,\omega)&=-  \frac{ \left(1-\frac{i\sigma(\omega)}{\varepsilon_0 \omega}\right) \beta }{\mu_0 c}\frac{\partial  \tilde{\phi}(r,\omega)}{\partial r}=\\
&=  \left(1-\frac{i\sigma(\omega)}{\varepsilon_0 \omega}\right)\frac{q_0 \beta c }{2\pi }\bar{\eta}C_0^{\sigma}(\omega)K_1\left(\bar{\eta}r\right)  
\end{split}
\end{equation}

\subsection{Surface Impedance of a cylindrical pipe}
By using the definition of $Z_s$ (Eq.\ref{eq:Z_s}), we get the expression of the surface impedance  on the wall of the cylindrical pipe:
\begin{equation}
\label{Z_s}
Z_{s}(\omega)
=\frac{\tilde{E}_z(b,\omega)}{\tilde{H}_\varphi(b,\omega)}=Z_0\frac{\sqrt{-\frac{1}{\beta^2\gamma^2}-i\frac{\sigma(\omega)}{\varepsilon_0\omega}}}{\left(1-\frac{i\sigma(\omega)}{\varepsilon_0 \omega}\right)}\frac{K_0\left(\bar{
 \eta}b\right)}{K_1\left(\bar{
 \eta}b\right)}
\end{equation}
It's worth noting that when $\gamma \rightarrow\infty$ and with high conductivity, $Z_s$ tends to the surface impedance of a conductor lighted by a plane wave.\\
The continuity of  the axial electric and azimuthal magnetic fields on the pipe wall $r=b$ determine the parameters $B^\sigma_0$ and $C^\sigma_0$ uniquely : 
\begin{equation}
\label{coeff}
\begin{split}
    &C^\sigma_0(\omega) =\frac{I_0\left(\frac{\omega b}{\gamma\beta c}\right)}{\beta^2 \gamma^2\left(\frac{1}{\beta^2 \gamma^2}+i\frac{\sigma(\omega)}{\varepsilon_0 \omega}\right)K_0\left(\bar{\eta}b\right)}B^\sigma_0(\omega)\\
&B^\sigma_0(\omega)=\frac{ i\beta^2 \gamma^2 c}{ \omega b  I^2_0\left(\frac{\omega b}{\gamma\beta c}\right) \left[\frac{Z_0}{Z_s}+i\frac{\beta\gamma I_1\left(\frac{\omega b}{\gamma\beta c}\right)}{ I_0\left(\frac{\omega b}{\gamma\beta c}\right)}\right]}
\end{split}
\end{equation}
where we have considered Eq. \ref{Z_s} for the surface impedance $Z_s$.
\subsection{Longitudinal Coupling Impedance}
\begin{figure}[!h]
     \centering
     \begin{subfigure}[b]{0.9\columnwidth}
         \centering
\includegraphics[width=\textwidth]{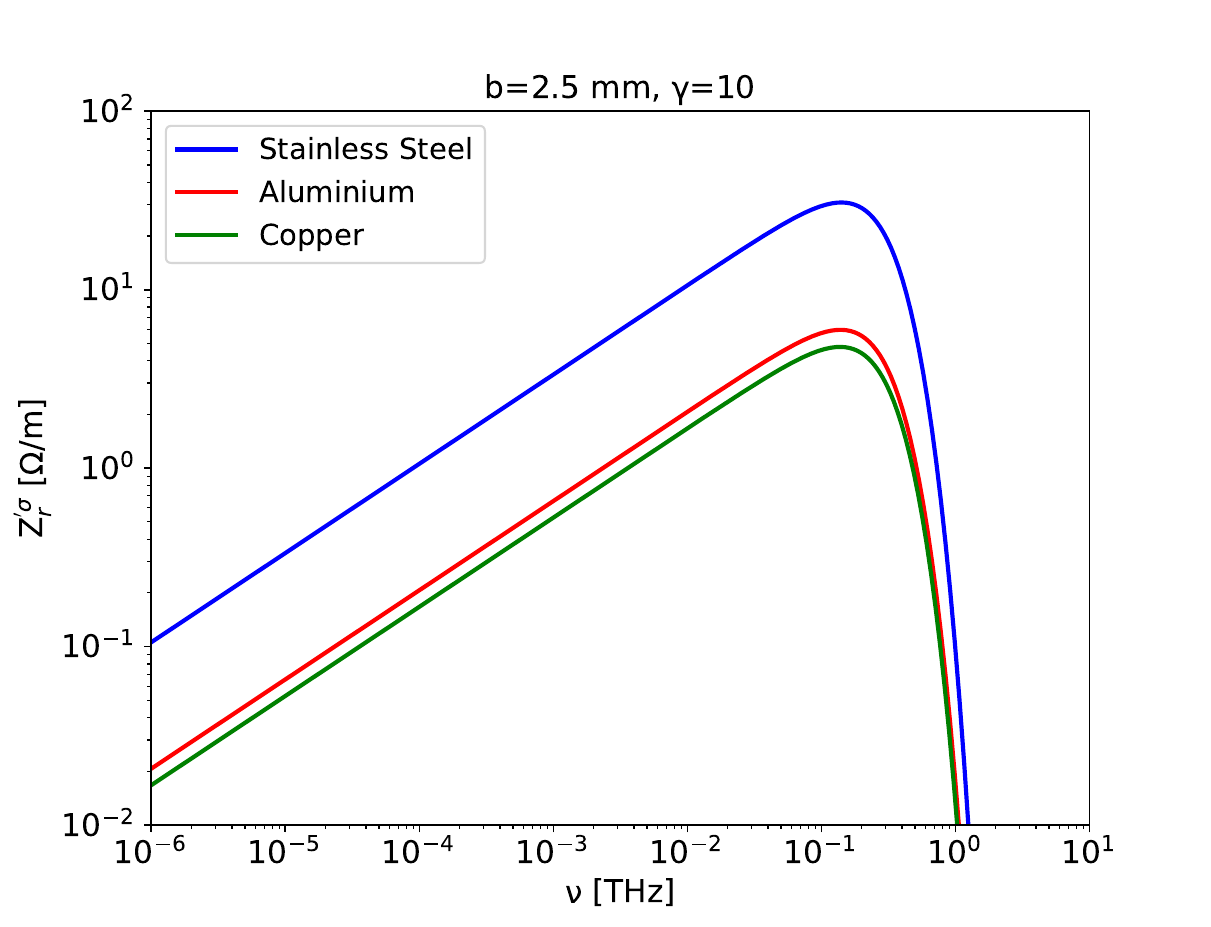}
    \caption{\label{ZrDC100}}
     \end{subfigure}
     \vfill
     \begin{subfigure}[b]{0.9\columnwidth}
         \centering
\includegraphics[width=\textwidth]{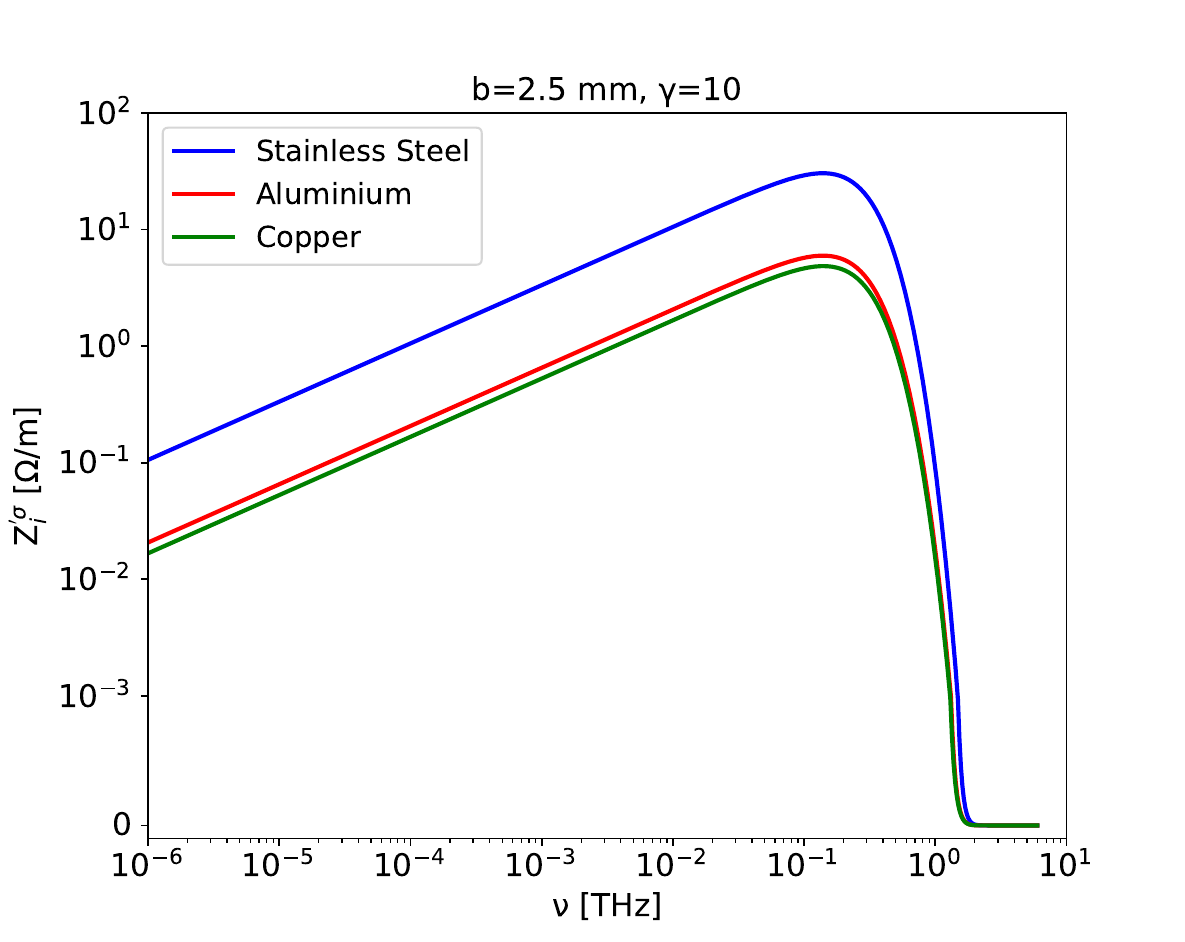}
\caption{\label{ZiDC100}}
     \end{subfigure}
        \caption{\label{ZDC100}   Real (\textbf{a}) and imaginary (\textbf{b}) part of coupling impedance $Z^\sigma$ vs frequency for different materials at normal condition, assuming DC electric conductivity  ($b=2.5\ mm$, $\gamma=10$).}
\end{figure}

By using Eq. \ref{coeff} in Eq. \ref{eq:impedance_sigma}, we obtain the longitudinal beam impedance per unit length:
\begin{equation}
\label{beamimped}
    Z'^\sigma(\omega)=\frac{dZ^\sigma(\omega)}{ds}=\frac{ Z_0}{2\pi b  I^2_0\left(\frac{\omega b}{\gamma\beta c}\right) \left[\frac{Z_0}{Z_{s}}+\frac{i\beta\gamma I_1\left(\frac{\omega b}{\gamma\beta c}\right)}{ I_0\left(\frac{\omega b}{\gamma\beta c}\right)}\right]}
\end{equation} 
It's worth of note that for $\gamma\rightarrow\infty$, Eq. \ref{beamimped} tends to:
\begin{equation}
\label{comparison}
        Z'^\sigma(\omega)\simeq\frac{ Z_0}{2\pi b  \left[\frac{Z_0}{Z_s}+\frac{i\omega b}{ 2c}\right]}
\end{equation}
that was found in Ref. \cite{Chao_1993} for a beam with the shape of a thin ring propagating with velocity $c$ in a cylindrical pipe.
\\
Eq. \ref{beamimped} has a general validity; the longitudinal impedance depends on the energy $\gamma$, the conductivity $\sigma(\omega)$ and the radius of the pipe $b$.
The Bessel function $I_0$ at the denominator shows that even for a point charge, the impedance frequency spectrum does not extend much beyond the frequency $\omega^*= \gamma\beta c/b$ where there is exponential decay. In fact, due to the relativistic contraction of the primary fields, the point charge has an effective excitation length $b/\beta\gamma$ on the conducting walls. Noticeably, the exponential drop does not exist if $\gamma \to \infty$.
\\
If $\omega^*\tau<<1$, the impedance spectrum of the point charge essentially falls below the frequency $1/\tau$, thus it is sufficient to consider a DC conductivity (Eq. \ref{sigma0}) in the expression of the surface impedance, since $\omega\tau<<1$ is always verified. \\ Conversely, if $\omega^* \tau \gtrsim1$, one has to consider the general AC conductivity (Eq. \ref{con1}). 
For instance, in the case of copper, where $\sigma_0 \simeq 5.7\times10^7\ \Omega^{-1} m^{-1}$ and $\tau \simeq 2.2 \times 10^{-14} s$, the DC condition is verified when $b/\beta\gamma >> 7 \times 10^{-6} m$. 
\\
In Figs. \ref{ZrDC100} and \ref{ZiDC100}, we show a case of DC conductivity, corresponding to $\gamma=10$ and $b=2.5\ mm$, for three different materials whose parameters are reported in Tab. \ref{tab1}.
\begin{table}
\caption{\label{tab1} Electric properties of different conductors}
\begin{center}
\begin{tabular}{||c c c c||} 
 \hline
 Metal & $\sigma_0  ( \times 10^7$ S/m) & $\tau$ (fs) & $\omega_p (\times 10^{16}$rad/s) \\ [0.5ex] 
 \hline\hline
 Cu & 5.7 & $\simeq 22$ & $\simeq1.7$ \\ 
 \hline
 Al & $3.7$ & $\simeq 7$ & $\simeq2.4$ \\
 \hline
 Steel  & $0.14$ & $\simeq 0.6$ & $\simeq1.6$ \\
  [1ex] 
 \hline
\end{tabular}
\end{center}
\end{table}
The real and imaginary parts of the coupling impedance $Z'^\sigma$ versus frequency are reported. 
\\
In the log-log plot, we note a large frequency region where the impedance grows with a constant power of the frequency, reaching a maximum before the exponential decay. In this case, the exponential drop is apparent at about $1\ THz$. As expected, the
low-frequency impedance scales as $\propto \sqrt{\omega/\sigma_0}$.
For the parameters considered in Figs. \ref{ZrDC100} and \ref{ZiDC100}, the high frequency coupling impedance can be approximated as:
\begin{equation}
Z'^\sigma(\omega)\simeq\frac{ Z_s}{2\pi b  I^2_0\left(\frac{\omega b}{\gamma\beta c}\right) }\simeq \frac{\sqrt{i\varepsilon_0 \omega}}{ b\sqrt{\sigma_0}  }\left(\frac{\omega b}{\gamma\beta c}\right)e^{-\frac{2\omega b}{\gamma\beta c}}
\end{equation}
The above expression corresponds to a maximum obtained for the following value of frequency:
\begin{equation}
\label{ommax}
    \omega_{max}=\frac{3 \gamma\beta c}{4 b}=\frac{3}{4}\omega*
\end{equation}
which does not depend on conductivity, as evident from Fig. \ref{ZrDC100}, approximately described by Eq. \ref{ommax}.
For larger energies/smaller pipe radii, in addition to a maximum, the beam impedance can acquire a resonance peak, which is well described by the poles of the $Z'^\sigma$ function, treating $\omega$ as a complex variable. The analytical derivation of the resonance frequency has been obtained in Ref. \cite{bane1996short}, in the case $\gamma\rightarrow\infty$.
\begin{figure}[!h]
     \centering
     \begin{subfigure}[b]{0.9\columnwidth}
         \centering
\includegraphics[width=\textwidth]{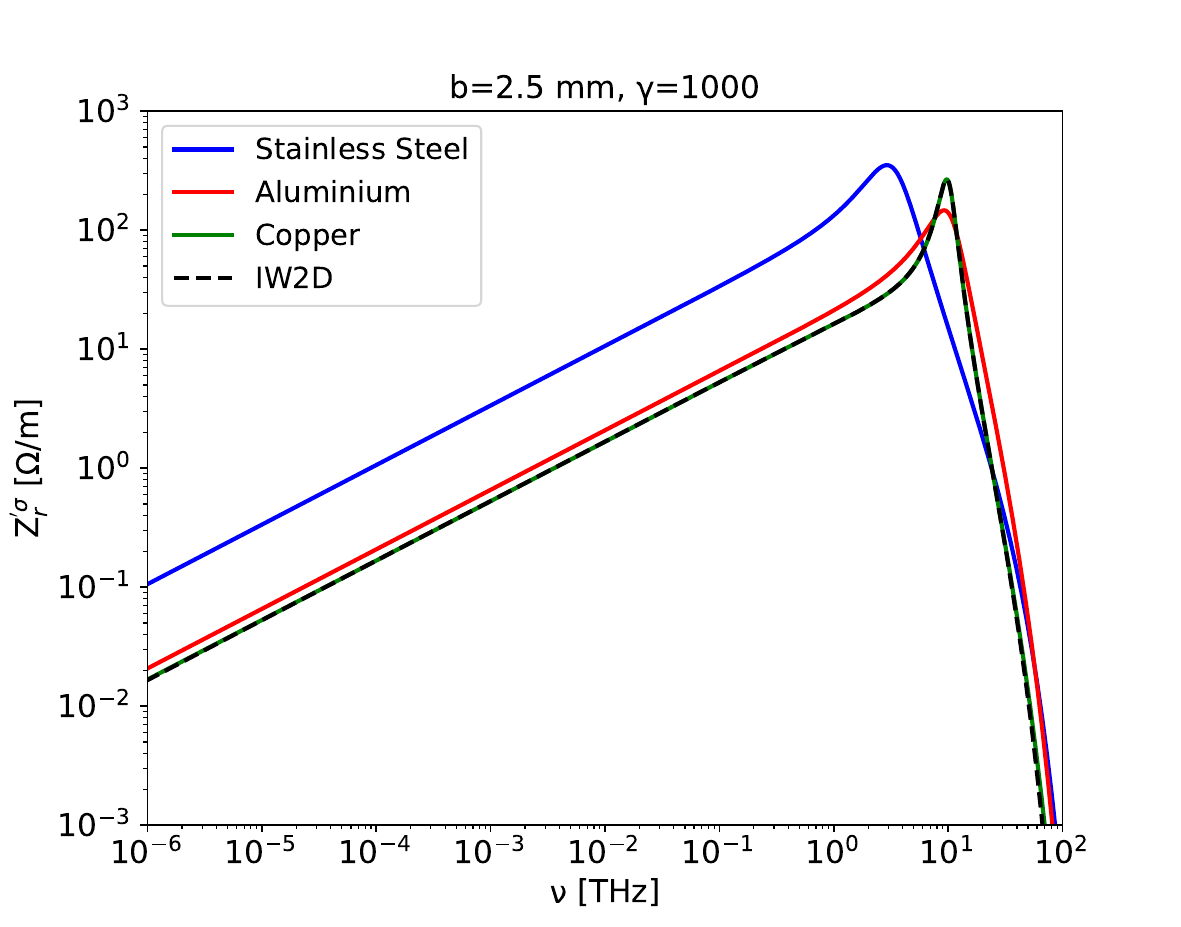}
    \caption{\label{ZrAC100}}
     \end{subfigure}
     \vfill
     \begin{subfigure}[b]{0.9\columnwidth}
         \centering
\includegraphics[width=\textwidth]{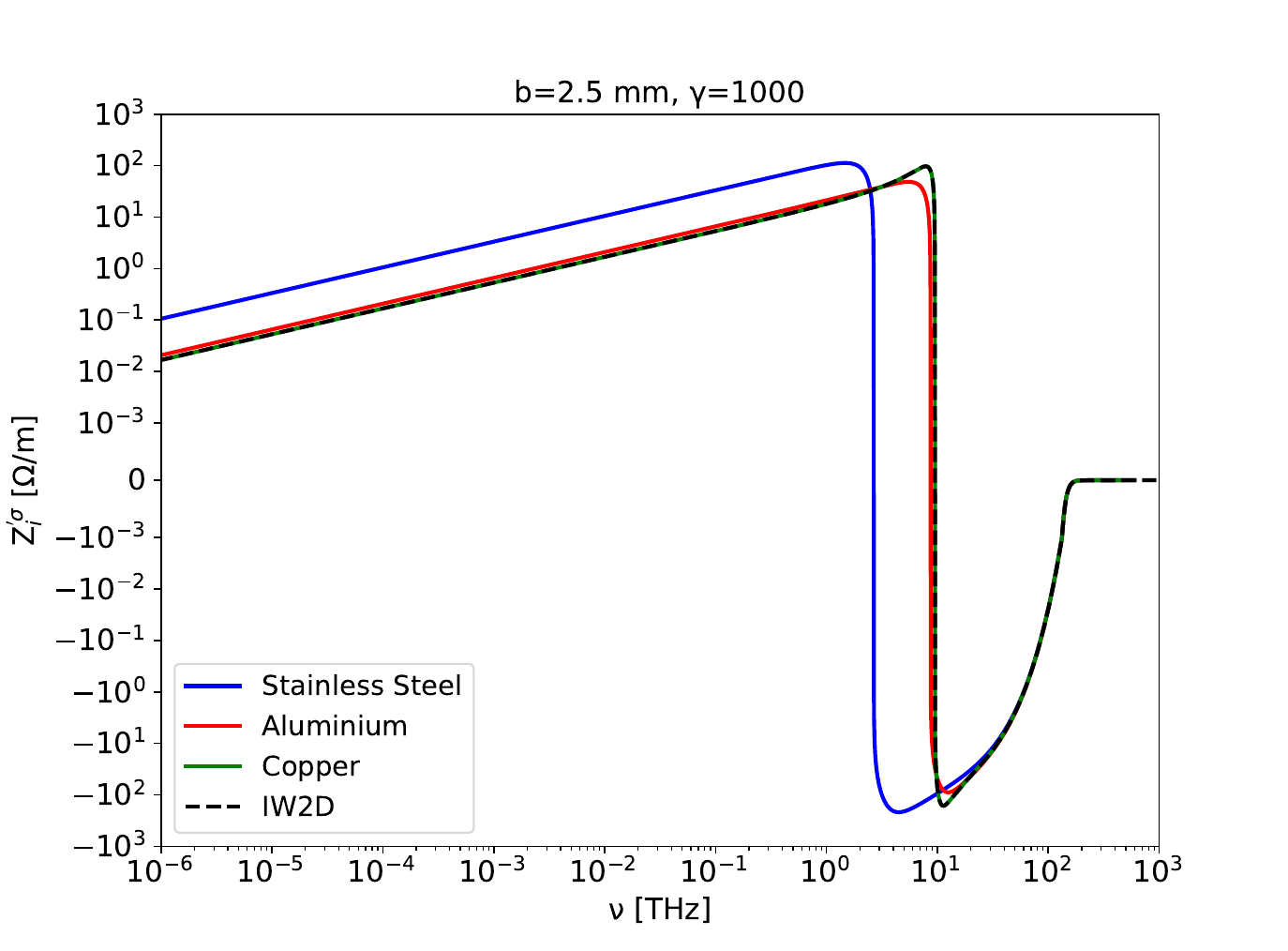}
         \caption{\label{ZiAC100}}
     \end{subfigure}
        \caption{\label{ZAC100}    Real (\textbf{a}) and imaginary (\textbf{b}) part of coupling impedance $Z'^\sigma$ vs frequency for different materials at normal condition, assuming AC electric conductivity  ($b=2.5\ mm$, $\gamma=1000$). For copper, it is also provided a direct comparison with the coupling impedance calculated through the IW2D code.}
\end{figure}
In general, the poles of the longitudinal beam impedance are determined by  the zeros of the following equation: 
\begin{equation}
\label{poles}
\frac{Z_0}{Z_{s}}+\frac{i\beta\gamma I_1\left(\frac{\omega b}{\gamma\beta c}\right)}{ I_0\left(\frac{\omega b}{\gamma\beta c}\right)}=0   
\end{equation}
\\
In Figs. \ref{ZrAC100} and \ref{ZiAC100}, we show the behavior of the longitudinal impedance for the same materials as in Figs. \ref{ZrDC100} and \ref{ZiDC100}, but in the AC regime of conductivity. The plots are similar, apart from a clear, sharp resonance appearing at high frequency.
If $\gamma\rightarrow\infty$, the resonance frequency can be analytically found as:
\begin{equation}
\label{omegaAC}
    \omega_{res}\simeq\left(\frac{4 c^2 \sigma_0}{\varepsilon_0 b^2 \tau}\right)^{1/4}=\sqrt{\frac{2c \omega_p}{b}}
\end{equation}
which coincides with the expression reported in Ref. \cite{bane1996short}.
It is easy to verify that the above expression fairly agrees with the results shown in Fig. \ref{ZrAC100} and \ref{ZiAC100}. An analytical expression for the resonance frequency can be found in the high energy limit ($\gamma>>1$) without any assumption on the value of the $\omega^\star\tau$ parameter.
This expression, long and cumbersome, is reported in Appendix \ref{resfreqapp} for the sake of completeness, and correctly tends to Eq. \ref{omegaAC} for $\gamma\rightarrow\infty$.\\
Furthermore, for copper, we also provided a direct comparison with the coupling impedance calculated through the IW2D (ImpedanceWake 2D \cite{Mounet_thesis}) code.
ImpedanceWake2D  computes the 
beam coupling impedances and wake functions in a multilayer axisymmetric or flat structure of infinite length. 
The code relies on the computation of the electromagnetic fields created by a point-charge beam traveling at any speed.
To compute the wake functions, an algorithm based on a Filon's kind
of method is used, instead of an FFT.  The codes automatically refine the frequency sampling until a
prescribed accuracy on the wake function is reached. \\
The correspondence between the analytical results and the code results is perfect, demonstrating the robustness of the formulas provided in this work.
\subsection{Loss factor}
 The energy loss per unit length $k'_\infty$ calculated via Eq.\ref{k'} in the limit $\gamma \to \infty$ is :
\begin{equation}
\label{Chaores}
    k'_\infty=\frac{1}{2\pi \varepsilon_0 b^2} 
\end{equation}
which does not depend on the wall's conductivity \cite{Chao_1993, bane1996short}. 
\begin{figure}[!h]
    \centering
    \includegraphics[width=85 mm]{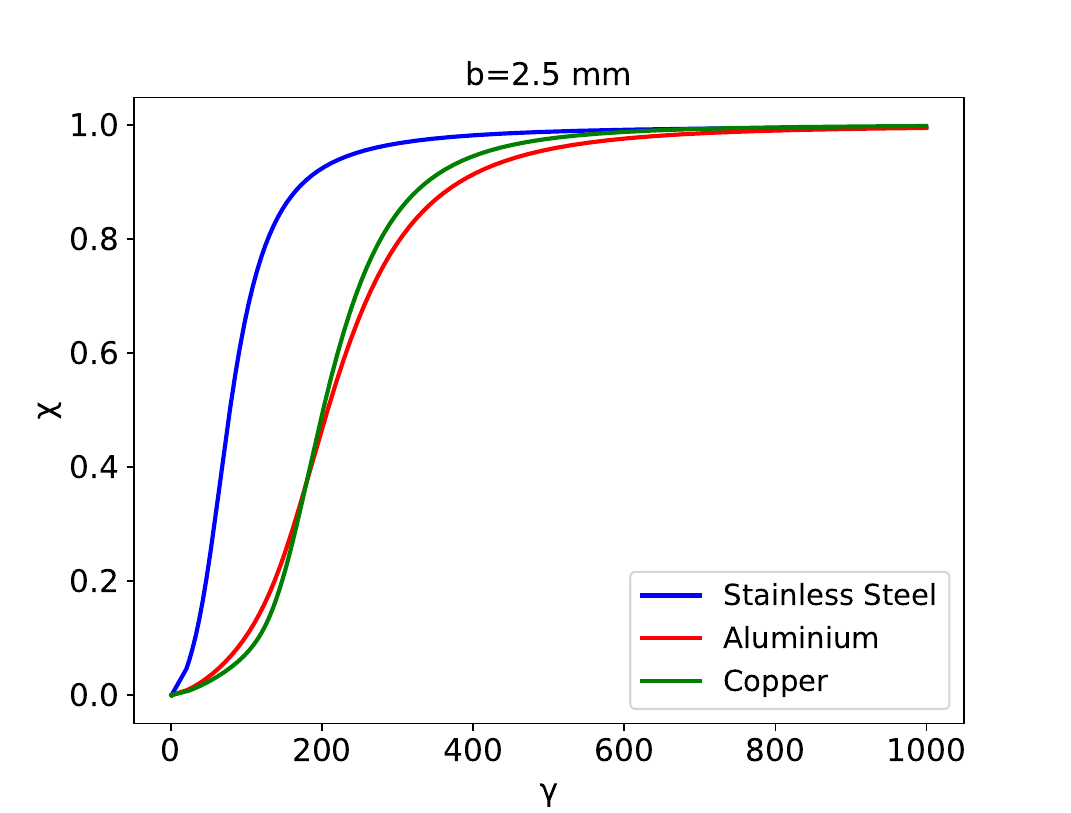}
    \caption{\label{lossgamma} Relative loss rate $\chi$ (between finite and infinite $\gamma$ theories, $\chi=1$ for infinite $\gamma$) versus $\gamma$, for the case of a cylindrical beam pipe with radius $b=2.5\ mm$.}
\end{figure}
For the real cases $\beta<1$ analyzed in the present paper, the loss factor is a function of energy, pipe's radius and conductivity. 
In the following, we calculate the normalized loss rate $\chi=k'/k'_\infty$, which is valuable to compare finite $\gamma$ results with the limit case of infinite $\gamma$.
\\
In Fig. \ref{lossgamma}, as an example, we show the relative loss $\chi$ versus the $\gamma$ parameter for the case of a cylindrical beam pipe with radius $b=2.5\ mm$ assuming the three materials of Tab. \ref{tab1}.
As one can note, there is a sharp transition between the range of validity of Eq. \ref{Chaores} and the range of parameters in which the finite $\gamma$ theory has to be applied. The transition is smoother for copper and aluminum, which have a $\sigma_0$ of the order of several times $10^7\ \Omega^{-1} m^{-1}$; with $b=2.5\ mm$ the asymptotic value is reached for $\gamma >500$.



\subsection{Wake potential}

By the inverse Fourier transform  Eq.\ref{wake_from_impedance}, we can derive the wake potential $w'_z(z)$. In Fig.4, we show the wake potential  calculated for a copper beam pipe (radius $2.5\ mm$) versus the $\gamma$ parameter.
As expected, the wake potential at the position $z=0$ is continuous for any value of $\gamma$, and the curve tends to be discontinuous for $\gamma \to \infty$ fulfilling the beam loading theorem (Eq. \ref{eq:beam_loading}).
The wake potential behind the leading charge ($z<0$) oscillates with a frequency which is well described by Eq. \ref{omegaAC} for high values of $\gamma$.
\begin{figure}[!h]
    \centering
\includegraphics[width=85 mm]{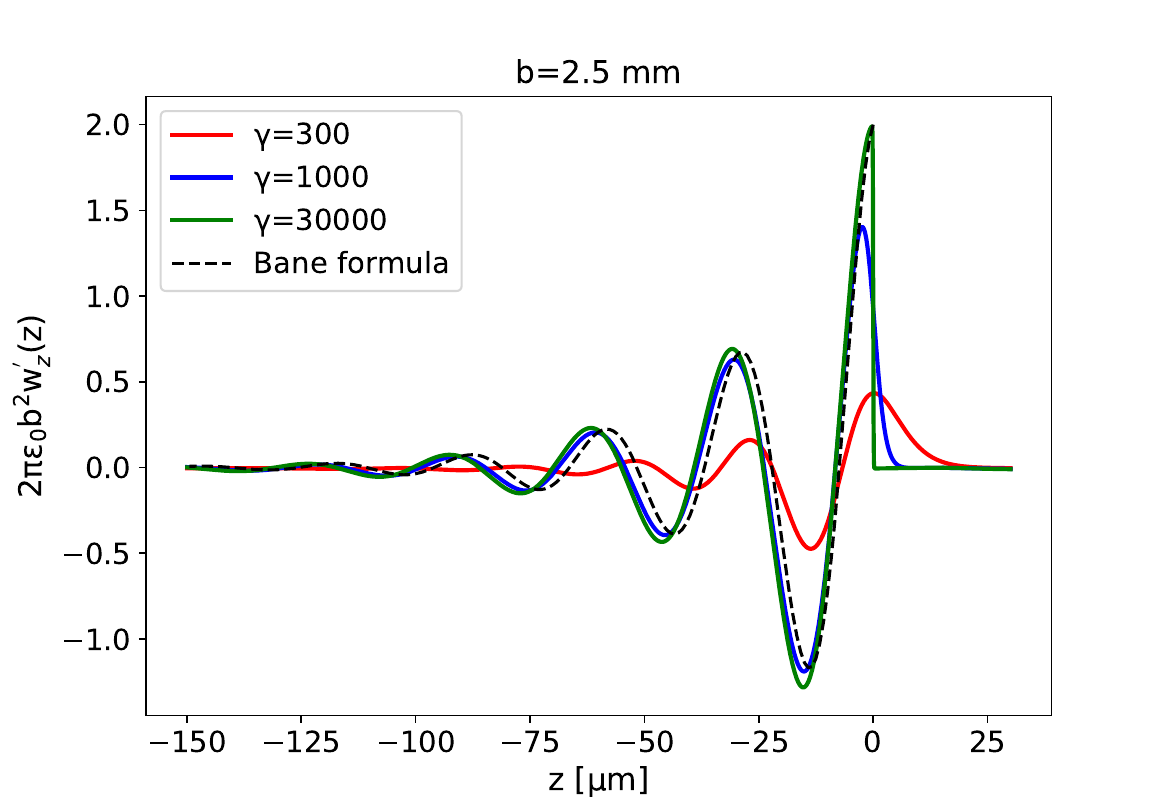}
    \caption{\label{WWK} Examples of wake potential calculated for a beam pipe made of copper ($b=2.5\ mm$) versus the $\gamma$ parameter. The chosen normalization shows consistency with the beam loading theorem.}
\end{figure}
Furthermore, the dashed line represents the result obtained through the Bane's formula in Ref. \cite{bane1996short}, which was obtained for infinite $\gamma$ and for the short-range wakefield neglecting diffusion terms, and that is reported here below:
\begin{equation}
\label{baneform}
w'_z(z\leq0)=2w'_z(0)e^{\frac{z}{4c \tau}}\cos{\left(\frac{\omega_{res}z}{c}\right)}
\end{equation}
The analytical expression in Eq. \ref{baneform}  approximates the wakefield produced by high-energy particles reasonably well.\\
As far as our theory is concerned, this includes the diffusion term: in fact, in the wave equation for potentials, the diffusion term is represented by the term proportional to the first derivative in the radial coordinate. The Bessel functions, as also explained in Ref. \cite{bane1996short}, are the exact solution of the problem and naturally include the diffusion effects, well describing both the short- and long-range wakefields.
\section{Electromagnetic fields inside the pipe's walls: Anomalous regime}
For the calculation of the electromagnetic field in anomalous conductivity conditions inside the pipe walls, we consider the following wave equation:
\begin{equation}
\label{phieq1}
   \frac{\partial^2 \tilde{\phi}}{\partial r^2}+\frac{1}{r}\frac{\partial \tilde{\phi}}{\partial r}-\left(\frac{k_z^2}{\gamma^2}\tilde{\phi}+i\mu_0 k_z\beta c (\sigma\star\tilde{\phi})_p\right)=0
\end{equation}
where the conductivity  $\sigma=\sigma(k_r,\omega)$ is function of both the radial and longitudinal wavevectors, $k_r$ and $k_z=\omega/\beta c$ respectively. The convolution $(\sigma\star\tilde{\phi})_p$ is intended to be applied in the radial plane, with respect to the coordinate $r$, and is dependent on the parameter $p$, as shown in Appendix \ref{reuterapp}. Specifically, the $p$-index is the probability of specular reflection of the charge carriers on the pipe surface. 
For the case of perfect specular reflection, $p=1$, while for completely diffuse reflection, $p=0$. For specular reflection, $(\sigma\star\tilde{\phi})_p$ is defined as (see Eq. \ref{convgenrad}):
\begin{equation}
\label{sconv1}
\begin{split}
&(\sigma\star\tilde{\phi})_{p=1}=\frac{3 n_0  e^2 }{4 m v_F } \int_0^{\pi/2} d\theta\int_0^{\infty}dr'\sin^3{\theta} \sec{\theta}\times\\
&\times\tilde{\phi}(r',\omega) e^{\left(\frac{(1+i \omega\tau)\sec{\theta}}{ v_F \tau} -\frac{i\omega \tan{\theta}}{\beta c }\right)(r-r')  }
\end{split}
\end{equation}
while, for $p=0$:
\begin{equation}
\label{sconv0}
\begin{split}
&(\sigma\star\tilde{\phi})_{p=0}=\frac{3 n_0  e^2 }{4 m v_F } \int_0^{\pi/2} d\theta\int_b^{\infty}dr'\sin^3{\theta} \sec{\theta}\times\\
&\times\tilde{\phi}(r',\omega) e^{\left(\frac{(1+i \omega\tau)\sec{\theta}}{ v_F \tau} -\frac{i\omega \tan{\theta}}{\beta c }\right)(r-r')  }
\end{split}
\end{equation}
The solution of Eq. \ref{phieq1} is rather complex.
For the case $p=1$, one can proceed in two ways. The first makes use of the similarity approach, seeking a solution analogue to that found in Sec. \ref{sec1}, while exploiting the properties of the Hankel-transform (see Appendix \ref{p1hank} for more details).\\ 
Otherwise, the wave equation for $p=1$ can be found by following an analogue procedure by Reuter and Sondheimer, as described in Appendix \ref{rsappro}. Such a procedure is based on the equivalence principle: the specular reflection of the carriers is justified by fictitious mirror charges placed in the space $r<b$, which, as field sources for the region $r>b$, behave equivalently to surface electric currents. In this picture, the field is decaying on the scale of the skin depth $\delta$ on both sides of the surface. This approach presents no issues for planar geometry, as demonstrated by Reuter and Sondheimer, where the half-space on one side of the surface is semi-infinite, as is the space on the opposite side. For cylindrical symmetry, additional complications may arise if the conductivity is not such that $\delta<<b$ in the region $r>b$. Therefore, a Reuter/Sondheimer-like approach must be intended for cylindrical symmetry only to be valid for good conductors.\\

\subsection{Surface Impedance in the anomalous regime}
For $p=0$ the solution can be found by generalizing the approach found in Ref. \cite{dingle1953anomalous}, originally developed for a planar surface, to cylindrical symmetry.
\\ In this section, we report the results found for $p=0$ (see Appendix \ref{p0dingapp}) and for $p=1$, the latter obtained through the Hankel-transform approach (see Appendix \ref{p1hank}).\\ Thus, for $r>b$, the electric field is found to be:
\begin{equation}
\label{efirest}
\tilde{E}_z(r,\omega)=\frac{i \omega}{\beta c}\frac{q_0}{2\pi \varepsilon_0}C^\sigma_0(\omega)\xi_p(r,\omega)
\end{equation}
where, again, we have defined a parameter $C_0^\sigma(\omega)$ that guarantees the continuity of the electric field at the vacuum-pipe interface.
The dependence of the field on $r$ is fully contained in $\tilde{\xi}_p(r,\omega)$.
In the Appendix \ref{reuterapp} it is demonstrated that:
\begin{equation}
\label{fif0}
\xi_0(r,\omega)=\int_0^{\infty} dk_r J_0(k_r r)k_r \tilde{\xi}_0(k_r,\omega)  
\end{equation}
where
\begin{equation}
\label{eqf}
\tilde{\xi}_0(k_r,\omega)=b \frac{ J_1\left(k_r b \right)}{k_r} e^{-\frac{k_r}{\pi}\int_0^\infty\frac{\log{\left[1+\frac{\frac{\omega^2}{\gamma^2 \beta^2 c^2}+i \mu_0 \omega \sigma(\kappa,\omega)}{\kappa^2}\right]}}{\kappa^2+k_r^2}d\kappa}
\end{equation}
While:
\begin{equation}
\label{fif1}
\xi_1(r,\omega)=\int_0^{\infty}d k_r \left(\frac{1}{\gamma^2}+\frac{i\beta^2\sigma(k_r,\omega)}{\varepsilon_0 \omega}\right)\frac{J_0\left(k_r r\right) k_r}{ k_r^2 +\bar{\eta}^2}
\end{equation}
The magnetic field in the resistive walls of the pipe is then found via Eq. \ref{HE}, considering the electric field in Eq. \ref{efirest}.
%
In the anomalous region, the surface impedance will differ according to the $p$-parameter that best fits the microscopic properties of the vacuum-pipe interface. Thus, we define a surface impedance dependent on $p$.
For $p=0$, as shown in Appendix \ref{p0dingapp}, it is possible to obtain:
\begin{equation}
\label{Zs0}
    Z_{s,0}\simeq
    -\frac{i \pi Z_0 \omega  }{c\int_0^\infty\log{\left[1+\frac{\frac{\omega^2}{\gamma^2 \beta^2 c^2}+i \mu_0 \omega \sigma(k_r,\omega)}{k_r^2}\right]}dk_r}
\end{equation}
Instead, for $p=1$, it is found that (see Appendix \ref{p1hank}):
\begin{equation}
\label{Zs1}
    Z_{s,1}=\frac{i \omega Z_0}{  c}\frac{\int_0^{\infty}d k_r \left(\frac{1}{\gamma^2}+\frac{i\beta^2\sigma(k_r,\omega)}{\varepsilon_0 \omega}\right)\frac{J_0\left(k_r b\right) k_r}{ k_r^2 +\bar{\eta}^2}}{  \int_0^{\infty}d k_r \left(\beta^2-\frac{i\beta^2\sigma(k_r,\omega)}{\varepsilon_0 \omega}\right)\frac{J_1\left(k_r b\right) k^2_r}{ k_r^2 +\bar{\eta}^2}}
\end{equation}
For good conductors, the surface impedance for $p=1$ can also be simplified into (see Appendix \ref{rsappro}):
\begin{equation}
\label{impedsurp1}
    Z_{s,1}\simeq-\frac{2 i Z_0 \omega b }{c} \int_0^{\infty}\frac{k_r J^2_0\left(k_r b\right)  dk_r}{k_r^2+\frac{\omega^2}{\gamma^2 \beta^2 c^2}+i \mu_0 \omega \sigma(k_r,\omega)}
\end{equation}
The above expressions of surface impedance are not only an extension of the Reuter theory to a cylindrical geometry, but also take into account the features of the primary fields, which are generated by a point charge traveling on the axis with speed $\beta c$. It is worth noting that Eq. \ref{Zs1} reproduces the classical expression Eq. \ref{Z_s} in the limit of local response of the charge carriers (normal region of conductivity) and, for good normal conductors, both Eqs. \ref{Zs0} and \ref{Zs1} reproduce the classical results.

\subsection{Beam Impedance in the anomalous regime}
\begin{figure}[!h]
   \centering
    \includegraphics[width=85 mm]{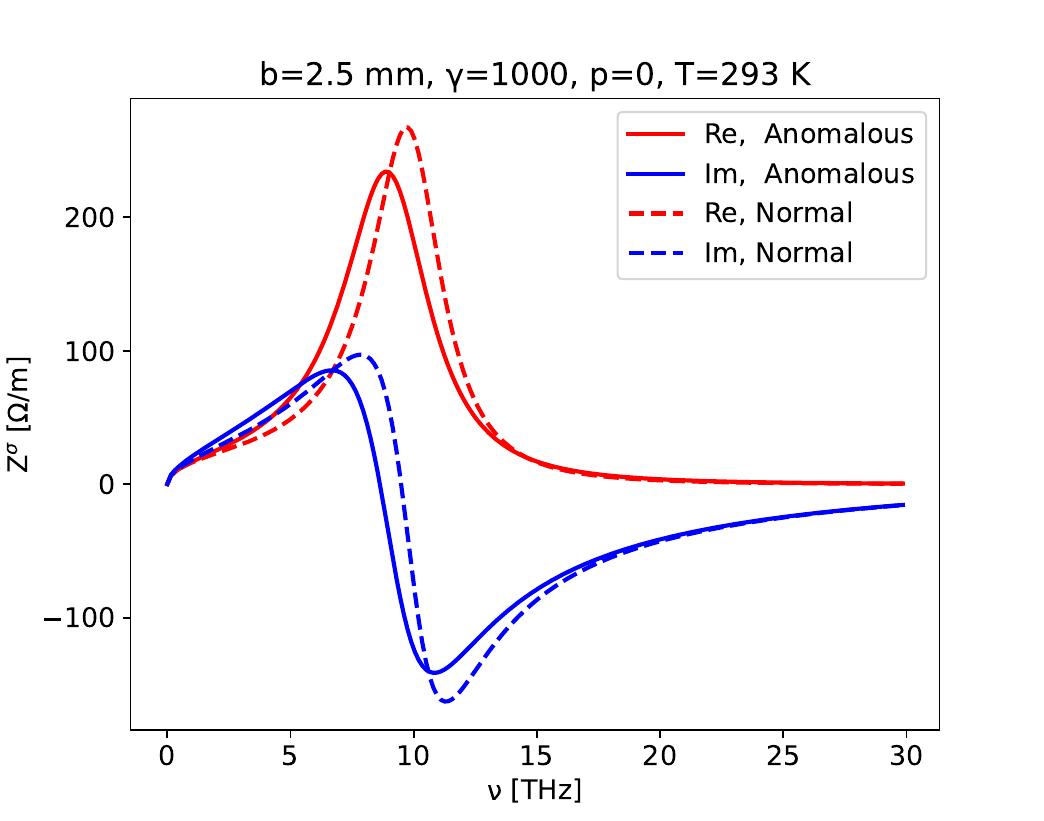}
   \caption{\label{normtemp} Longitudinal beam impedance for the case of a copper beam pipe of radius $b=2.5\ mm$, Lorentz factor $\gamma=1000$ and normal temperature. The continuous line considers the anomalous effect through Eq. \ref{Zs0}, while the dashed lines correspond to the same curves while neglecting the dependence of $\sigma$ upon $k_r$ in Eq. \ref{Zs0}.}
\end{figure}

We can obtain the beam impedance in the anomalous regime by using  Eqs. \ref{Zs0} and  \ref{Zs1} in Eq.\ref{beamimped}. Fig. \ref{normtemp} shows that in the case of diffuse carrier reflection ($p=0$), the longitudinal beam impedance is slightly affected by the anomalous conductivity even at room temperature. 
\begin{figure}[!h]
   \centering
    \includegraphics[width=85 mm]{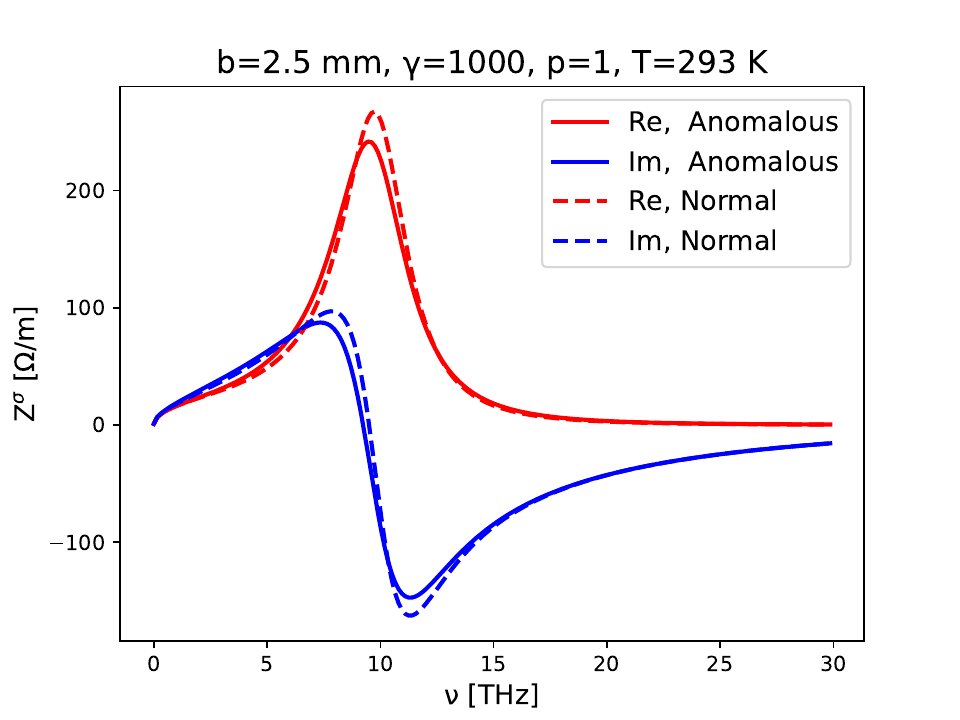}
   \caption{\label{normtemp1} Longitudinal beam impedance for the case of a copper beam pipe of radius $b=2.5\ mm$, Lorentz factor $\gamma=1000$ and normal temperature. The continuous line considers the anomalous effect through Eq. \ref{Zs1}, while the dashed lines correspond to the same curves while neglecting the dependence of $\sigma$ upon $k_r$ in Eq. \ref{impedsurp1}.}
\end{figure}
The same can be said regarding Fig. \ref{normtemp1}, for the case of specular carriers' reflection at the pipe surface ($p=1$). 
Figs. \ref{normtemp} and \ref{normtemp1} suggest that  Eq. \ref{Z_s} is an approximation even for conductive pipes at normal temperature, where the anomalous conductivity, i.e. the nonlocal response of the charge carriers, is not entirely negligible. Eq. \ref{Z_s} can be a reasonable approximation at normal temperature, depending on the precision sought in the frequency region of interest.
\\
To treat the cryogenic temperatures, it is customary to rescale the conductivity by a factor called $RRR=\sigma(4K)/\sigma(293K)$ (related to the residual resistivity at absolute zero). In the normal region, DC case, this is accomplished by multiplying $\sigma_0$ by $RRR$. In the AC case, the correct procedure is to rescale the relaxation time $\tau$ (it shall be noted that $\sigma_0\propto\tau$). A physical interpretation of the $RRR$ factor is related to the scattering of the carriers on the lattice, whose rate is reduced at lower temperatures. Therefore, in a rigorous sense,  the $RRR$ factor is applied to the relaxation time $\tau$ more than to the conductivity as a whole.
\begin{figure}[!h]
   \centering
    \includegraphics[width=85 mm]{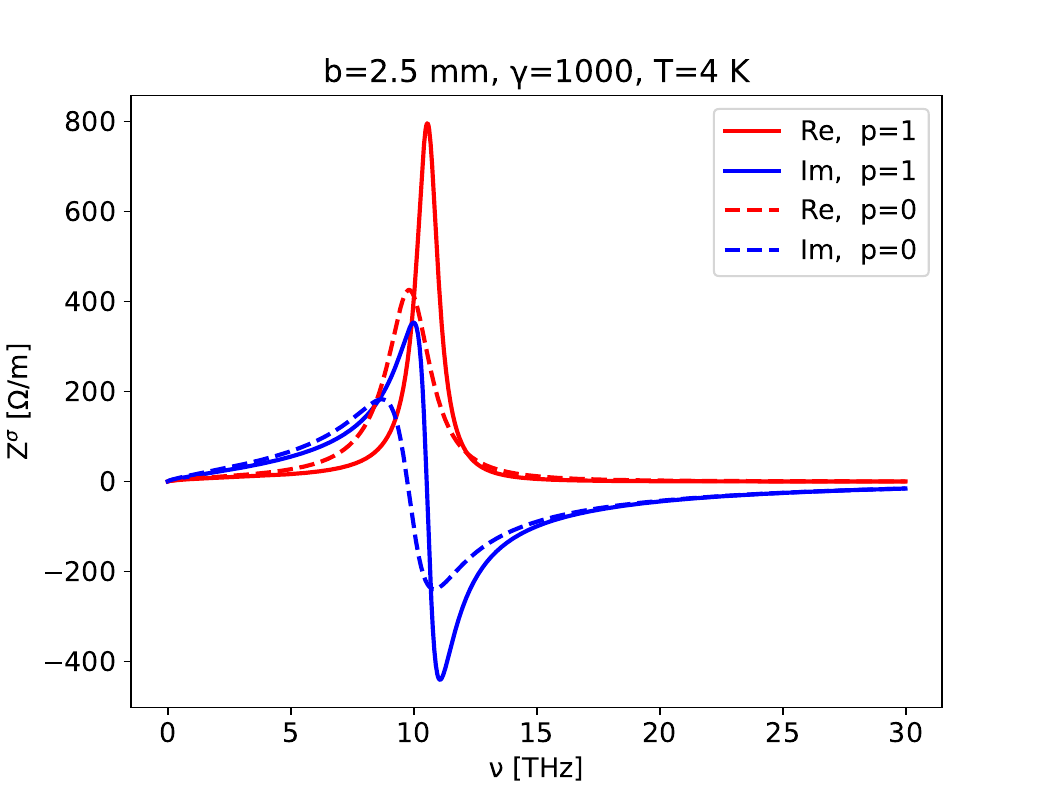}
   \caption{\label{cryotemp01} Longitudinal beam impedance for the case of a copper beam pipe of radius $b=2.5\ mm$, Lorentz factor $\gamma=1000$ and cryogenic temperature. Comparison between specular ($p=1$) and diffuse ($p=0$) reflection of the charge carriers on the surface of the pipe.}
\end{figure}\\
In the example of Fig. \ref{cryotemp01}, we have considered $RRR=100$. A comparison of two different dynamics of the charge carriers at the vacuum-metal interface is made for the longitudinal beam impedance.
However, in the literature \cite{dingle1953anomalous, Stupakov}, the most relevant cases  for real physical systems is reported to be $p=0$.\\
It is worth noting that applying Eq. \ref{Z_s} for the cryogenic case the beam impedance would show a peak two orders of magnitude higher and the resonance shifted towards higher frequencies. Therefore, for the cases where anomalous conductivity is not negligible (i.e. when the mean free path of the carriers is comparable or larger than the skin depth of the conductor), Eq. \ref{Z_s} is not applicable, while Eqs. \ref{Zs0}, \ref{Zs1}, or \ref{impedsurp1}, must be considered.
\subsection{Wake potential in the anomalous regime}
\begin{figure}[!h]
   \centering
    \includegraphics[width=85 mm]{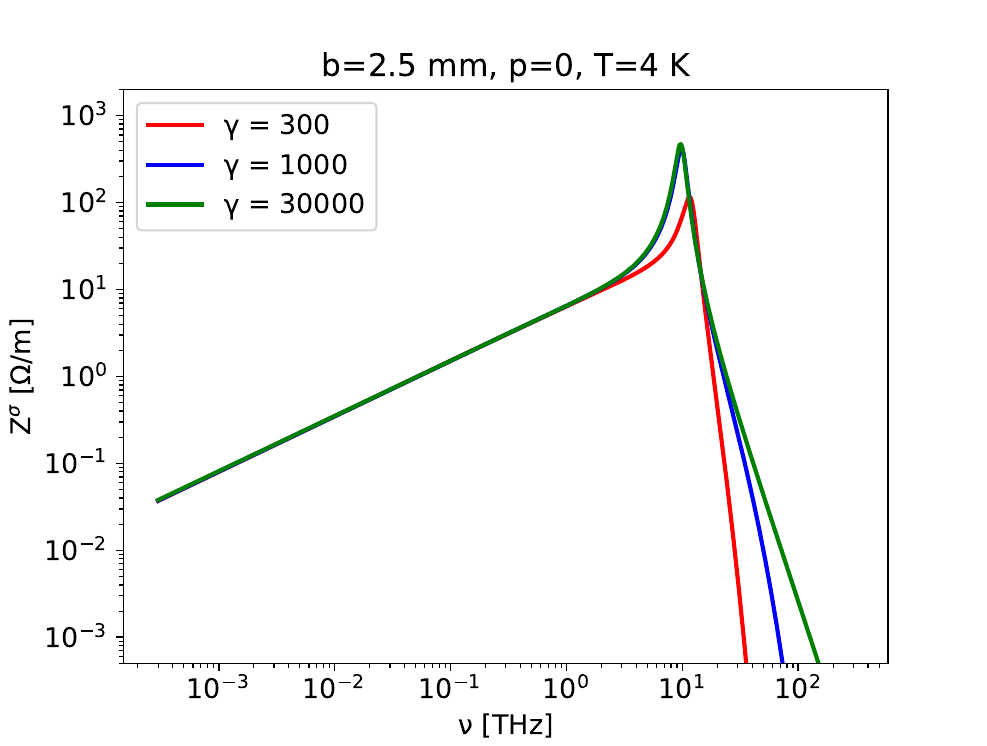}
  \caption{\label{p0difen} Longitudinal beam impedance for the case of a copper pipe of radius $b=2.5\ mm$, for different values of $\gamma$. Cryogenic temperature is assumed ($RRR=100$), and diffuse ($p=0$) reflection of the carriers on the surface of the pipe is assumed.}
\end{figure}

As the last examples provided in this work, we show wake potential curves in the anomalous regime for different particle energies. The considered beam pipe is made of copper at cryogenic temperature, with a pipe radius of $2.5\ mm$ and assuming diffuse reflection of the carriers at the vacuum-pipe surface ($p=0$). Before showing the simulation results on wake potentials, the longitudinal beam impedance is reported in Fig. \ref{p0difen}.
\begin{figure}[!h]
   \centering
    \includegraphics[width=85 mm]{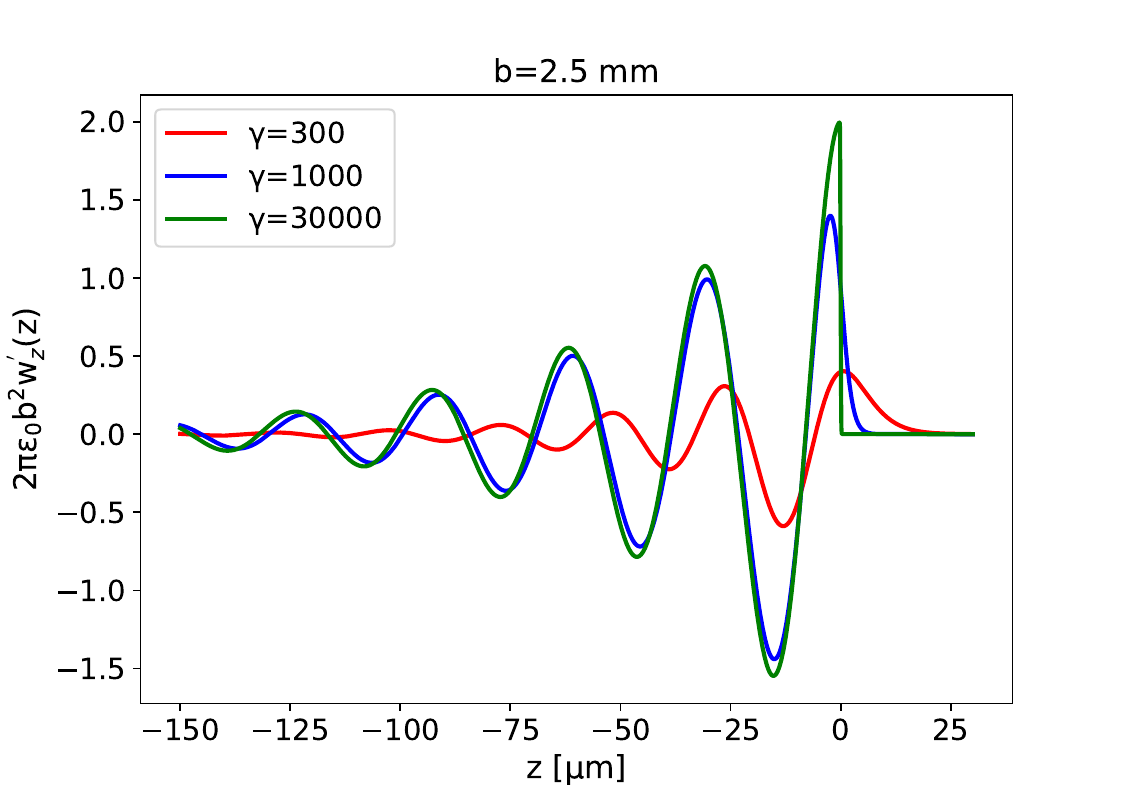}
  \caption{\label{wakeanom} Examples of wake potential calculated for a beam pipe made by copper ($b=2.5\ mm$) versus $z$ for three values of $\gamma$ parameter.  Cryogenic temperature is assumed ($RRR=100$) together with the carriers' diffuse reflection on the pipe's surface ($p=0$).}
\end{figure}
As expected, the
low-frequency impedance scales as $\propto \omega^{2/3}$ \cite{podobedov2009resistive}.\\
It is possible to notice that the beam impedance does not depend on the Lorentz factor at low frequency, as expected. However, the resonance frequency and the peak value of the impedance (therefore of the loss rate) sensitively depend on the particle energy for relatively low $\gamma$. 
The longitudinal wake potentials corresponding to Fig. \ref{p0difen} are reported in Fig. \ref{wakeanom}. The frequency of the wakefield is not significantly changed compared to the case studied at normal temperature (Fig. \ref{WWK}). However, the increase in electric conductivity and the narrowing of the resonance peak at cryogenic temperature determine a reduced damping of the oscillations.\\
Plots that are similar to Fig. \ref{wakeanom}, but valid in the extremely anomalous regime $|\mathcal{s}|>>1$, can be found in Ref. \cite{podobedov2009resistive}.
As in Ref. \cite{podobedov2009resistive}, the long-range wakefield corresponding to Fig. \ref{wakeanom} scales as $|z|^{-5/3}$. \\
Furthermore, it is worth emphasizing here, that $\gamma=30000$ is essentially equivalent to the case of infinite $\gamma$.

\section{Applications to bunched beams}
In this last section, we show simulation results for the wake potentials associated with monoenergetic particle bunches. 
\begin{figure}[!h]
     \centering
     \begin{subfigure}[b]{0.9\columnwidth}
         \centering
\includegraphics[width=\textwidth]{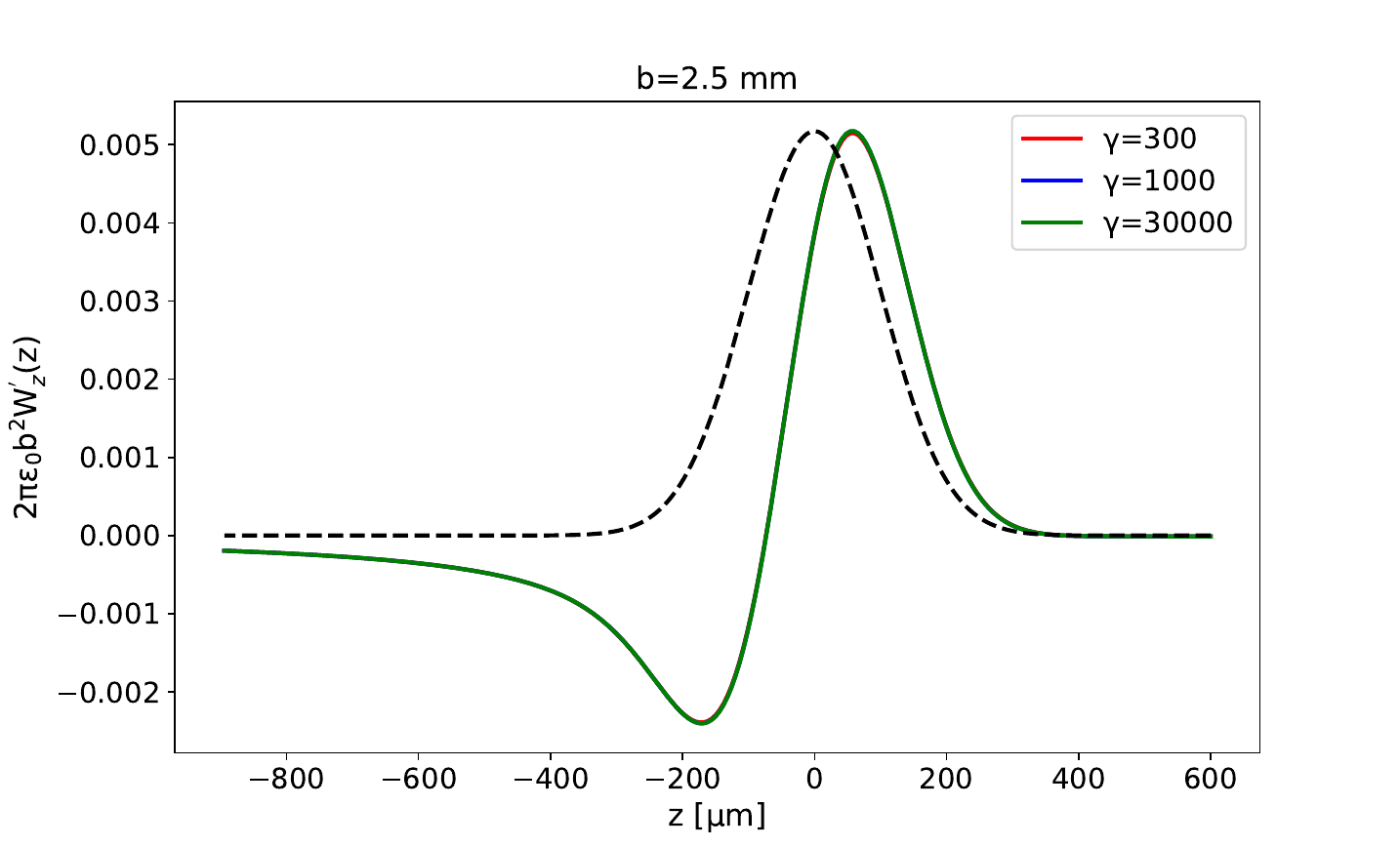}
\caption{\label{wakebunchnorm1}}
     \end{subfigure}
     \vfill
     \begin{subfigure}[b]{0.9\columnwidth}
         \centering
\includegraphics[width=\textwidth]{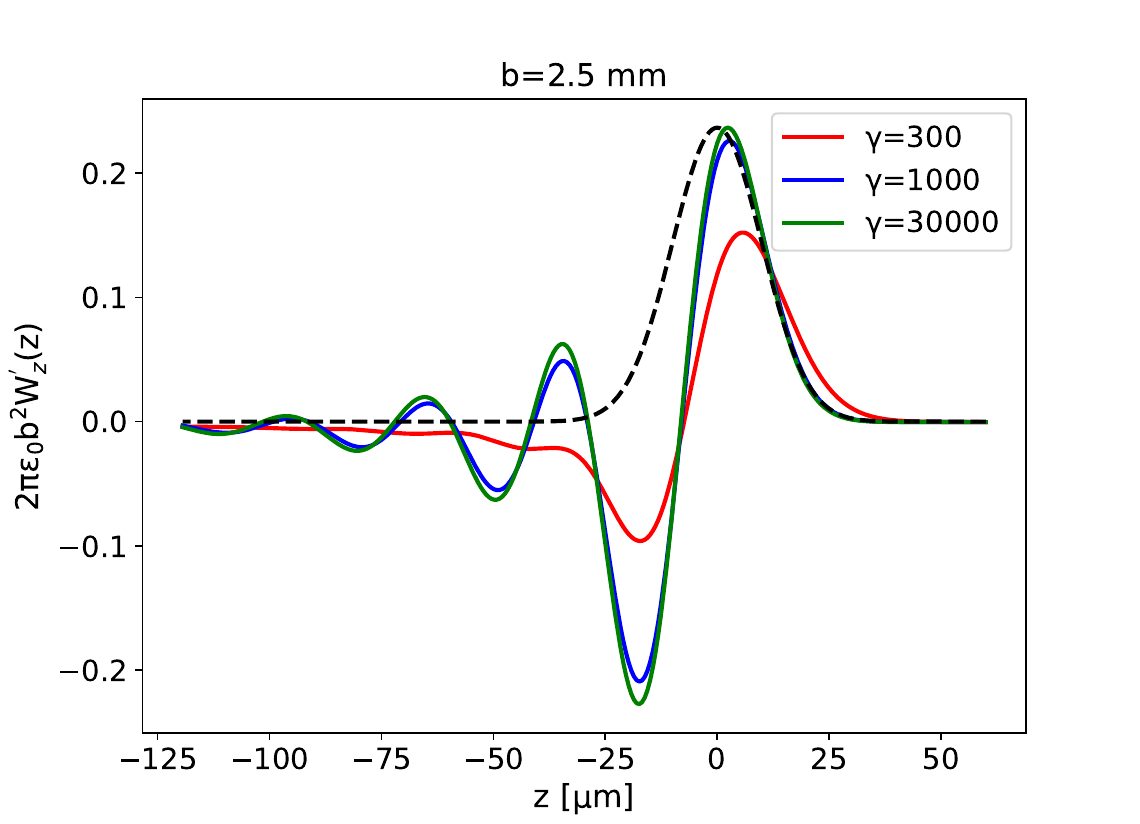}
\caption{\label{wakebunchnorm2}}
     \end{subfigure}
\caption{\label{WakeBunchNorm} Examples of wake potentials calculated for particle bunches propagating in a beam pipe made of copper ($b=2.5\ mm$) versus $z$ for three values of the $\gamma$ parameter, at normal temperature. Rms bunch length (\textbf{a}): $\sigma_z=100\ \mu m$, (\textbf{b}): $\sigma_z=10\ \mu m$). The dashed lines indicate the current profiles (arbitrarily scaled).}
\end{figure}
We use Eq. \ref{wakebunformula} to calculate the wakefield produced by bunches of different rms lengths, i.e. $\sigma_z=100\ \mu m$ and $\sigma_z=10\ \mu m$, where a gaussian shape it is assumed, therefore:
\begin{equation}
    \tilde{\lambda}(\omega)=\frac{Q_b}{ \beta c}e^{-\frac{\omega^2\sigma_z^2}{2\beta^2 c^2}}
\end{equation}
\begin{figure}[!h]
     \centering
     \begin{subfigure}[b]{0.9\columnwidth}
         \centering
\includegraphics[width=\textwidth]{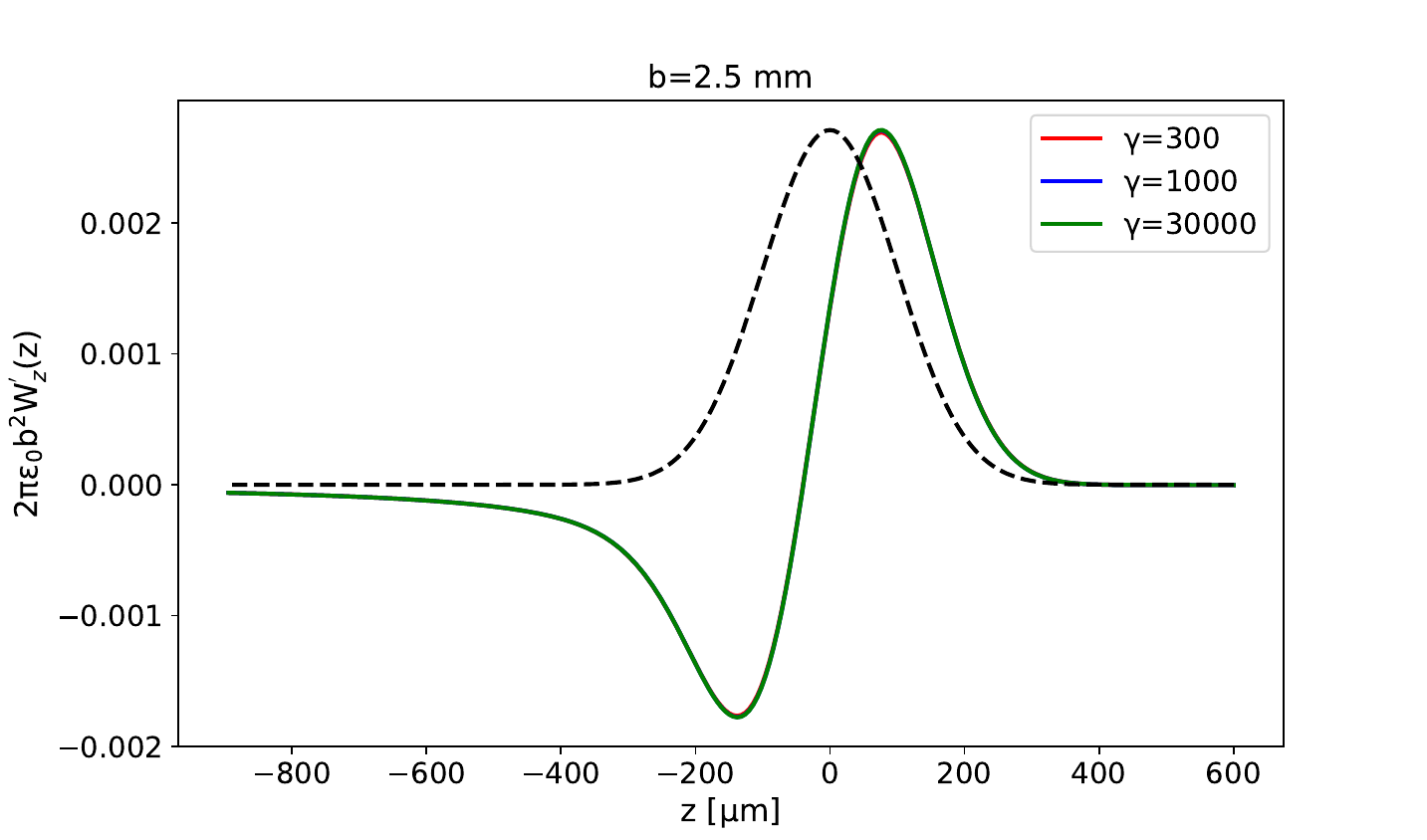}
\caption{\label{wakebunchcryo1}}
     \end{subfigure}
     \vfill
     \begin{subfigure}[b]{0.9\columnwidth}
         \centering
\includegraphics[width=\textwidth]{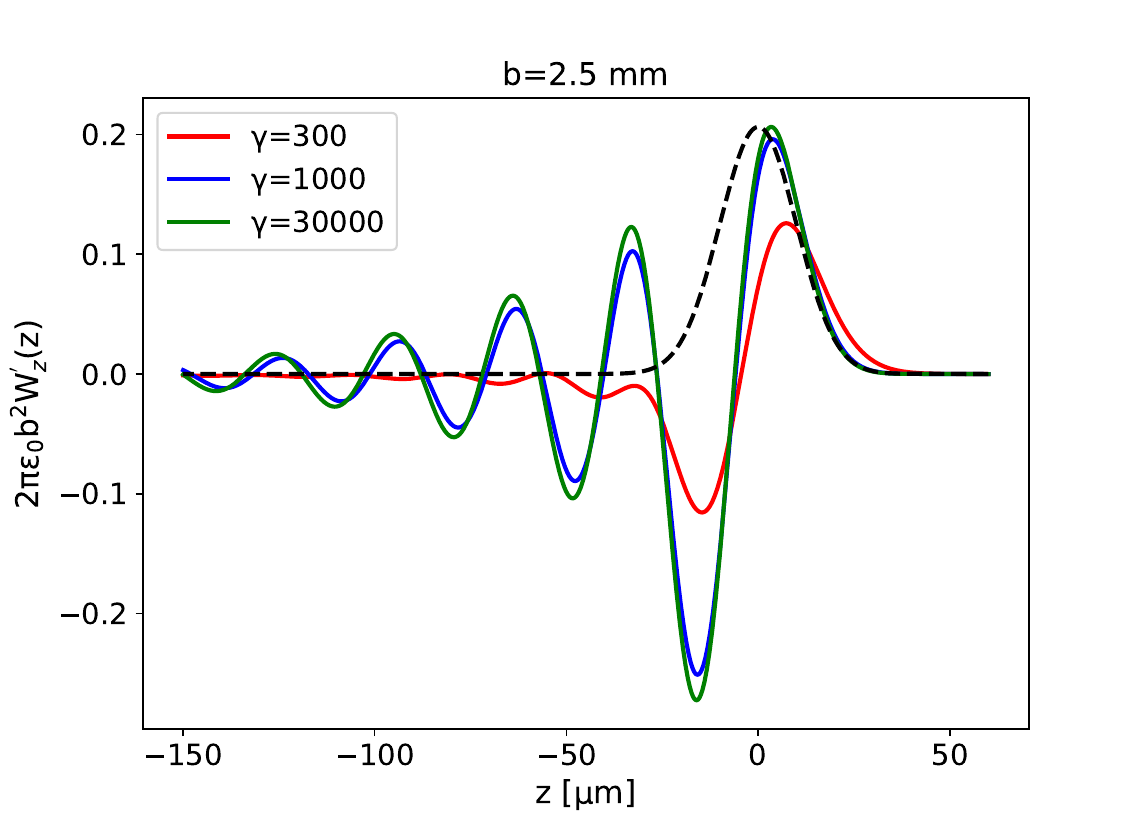}
\caption{\label{wakebunchcryo2}}
     \end{subfigure}
\caption{\label{WakeBunchCryo} Examples of wake potentials calculated for particle bunches propagating in a beam pipe made of copper ($b=2.5\ mm$) versus $z$ for three values of the $\gamma$ parameter, at cryogenic temperature. Rms bunch length (\textbf{a}): $\sigma_z=100\ \mu m$, (\textbf{b}): $\sigma_z=10\ \mu m$). The dashed lines indicate the current profiles (arbitrarily scaled).}
\end{figure}
Moreover, the calculation is performed for different beam energies and different regimes of conductivity, i.e. at normal temperature (Fig. \ref{WakeBunchNorm}) and cryogenic temperature (Fig. \ref{WakeBunchCryo}). It is apparent that, while there is no substantial difference in the wakefield of the longer bunch with different energy, for shorter bunches a high-frequency resonance is excited in both normal and anomalous regimes, which causes energy spread inside the beam and can affect the dynamics of trailing bunches.\\
The wakefield generated by relatively long bunches includes only the low-frequency "content" of the coupling impedance since the form factor $\tilde{\lambda}(\omega)$ cuts off all the frequencies such that $\omega>>\beta c/\sigma_z$. If $\omega^*\sigma_z/\beta c>>1$, $(\sigma_z>>b/\gamma)$, as for the cases studied in Figs. \ref{wakebunchnorm1} and \ref{wakebunchcryo1}, the  coupling impedance for a monoenergetic particle beam does not depend on the beam energy, as observed in the simulation results.\\
In the anomalous regime, there is less attenuation of the fields, that are ringing over a longer distance.
\\
Thus, the effects of conductivity become an important issue in designing Free-Electron Lasers (FELs) and, more in general, facilities that exploit ultrashort bunches; it modifies the charge distribution and the energy spread, affecting the bunch dynamics and directly influencing the overall performance and efficiency of such machines.\\
Figs. \ref{WakeBunchNorm} and \ref{WakeBunchCryo} are introduced for completeness, to discuss the wakefield produced by a bunch of particles rather than by a single particle, to highlight that the frequency cutoff imposed by a bunch can prevail over that of the single particle, and to better visualize the difference between normal and anomalous regimes at different beam energies. For the cases of relatively long bunches (low frequency excitation) and relatively large $\gamma$, our results are equivalent to those found, e.g., in references \cite{Chao_1993, podobedov2009resistive}, hence the similarity of our plots with those in the above-cited references.
\section{Conclusions}
This paper outlines a rigorous study of surface and coupling impedance in cylindrical symmetry. 
To this end, we adopted the convention to separate the space-charge from the resistive wall impedance as in Refs. \cite{Palumbo_Vaccaro} and \cite{stupakov2020resistive} but differently from Ref. \cite{Zimmermann_Oide}. \\
The theory we developed is valid for any value of particle energy, thus it can provide an important tool for the correct evaluation of impedance, loss rate and wakefields in both normal and anomalous conductivity regimes.\\
Our goal is unifying approximate theories published in the past (although valid in many practical cases) into a single coherent framework, also for clarifying their range of validity.\\
We have developed a new theoretical approach and derived a rigorous formula for the electric conductivity of a cylindrical pipe excited by a charged particle (Eq. \ref{gencond}, or \ref{convgenrad}), starting from the Boltzmann theory of a Fermi gas for the carriers. Additionally, we obtained new general formulas for the surface impedance of cylindrical pipes (Eq. \ref{Z_s}), which also led to a new impedance expression (Eq. \ref{beamimped} via Eq. \ref{Z_s}). A thorough discussion has been provided on the transition from low to high energy and from low to high frequency, governed by the dimensionless parameter 
$\omega b/\gamma\beta c$. The analysis includes a rigorous treatment of the anomalous regime (Eqs. \ref{phieq1}, \ref{sconv1}, and \ref{sconv0}), and presents analytical solutions for the electromagnetic field in the cylindrical pipe at cryogenic temperatures (Eqs. \ref{efirest}, \ref{fif0}, \ref{fif1}). Finally, we derive new and rigorous expressions for the anomalous surface impedance for both 
$p=0$ and $p=1$ cases (Eqs. \ref{Zs0} and \ref{Zs1}/\ref{impedsurp1}).
\\
To be more specific in the comparison with previous literature, our Eq. \ref{beamimped} is formally similar to the result found in Ref. \cite{stupakov2020resistive}, however, the surface impedance that we consider is not the same. In fact, in Eq. \ref{beamimped}, we consider the rigorously derived surface impedance for a cylindrical beam pipe in any regime of conductivity and particle energy. Differently, Ref. \cite{stupakov2020resistive} considers the surface impedance for an electromagnetic plane wave impinging on a normal flat conductor. It's important to highlight that the surface impedance for an electromagnetic wave impinging on a flat surface (Reuter and Sondheimer formula \cite{reuter1948theory}) is a good approximation for very large $\gamma$, but it may be an inconsistent choice within a theory that aims to be general for any particle energy.\\
It's worth recalling that neither Ref. \cite{Zimmermann_Oide} nor \cite{stupakov2020resistive} treat the case of anomalous conductivity, and other works that have treated the anomalous regime (see Refs. \cite{Stupakov,podobedov2009resistive})  have been developed for infinite energy and only for specific cases of the anomalous conductivity ($p=0$ \cite{Stupakov}, extremely anomalous region \cite{podobedov2009resistive}).\\
Indeed, the real behavior of any anomalous material is determined by a combination of the two cases $p=0$ and $p=1$ (both derived in this paper). We posed the foundations for a general treatment of the cryogenic pipes, necessary for a comparison with  measurements of the wakefields effects in the anomalous regime.\\
Beside its theoretical value, the work is relevant to particle accelerator design, with particular attention to machines based on cryogenic structures and/or moderate-energy ultra-short particle bunches.\\
For instance, we mention the case of moderate $\gamma$ values ($\lesssim 500$) and millimetric pipes in the case of FELs piloted by sub-picosecond bunches, as well as the case of low-energy accelerators as hadron accelerators at non-fully relativistic energy. 

\onecolumngrid
\appendix
\section{Theory of the electric conductivity via the Boltzmann equation}
\label{condusec}
The Boltzmann equation describing the transport of a non-relativistic electron under the action of an external electromagnetic field and in the presence of collisions with background particles (atoms, electrons, ions) is given by \cite{grosso2013solid}:
\begin{equation}
\label{Boltz1}
    \frac{\partial f}{\partial t}-\frac{e}{m}\left(\vec{E}+\vec{\mathcal{v}}\times\vec{B}\right)\cdot\vec{\nabla}_{\mathcal{v}}f+\mathcal{v}\cdot\vec{\nabla}_{\vec{r}}f=-\frac{f-f_0}{\tau}
\end{equation}
The function $f$ represents the phase space density of the charge carriers (here assumed to be electrons), $-e$ is the electron charge, and $m$ is the effective mass of the carriers in the conduction band. The velocity vector $\vec{\mathcal{v}}$ represents here the velocity of the carriers. The phase space density at the equilibrium is $f_0$. The relaxation time of the system is $\tau$.
For non-relativistic carriers, the interaction with the magnetic field can be neglected. Moreover,  by using the perturbative approach $f=f_0+f_1$ and transforming the Boltzmann equation from the $(\vec{r},t)$ domain into the Fourier domain $(\vec{k},\omega)$, it is possible to obtain an equation for the perturbative term
\begin{equation}
\label{Boltz2}
    i \omega f_1-\frac{e n_e}{m}\vec{E}\cdot\vec{\nabla}_{\vec{\mathcal{v}}}f_{FD}-i\vec{\mathcal{v}}\cdot\vec{k}f_1=-\frac{f_1}{\tau}
\end{equation}
which is solved by:
\begin{equation}
\label{Boltz30}
    f_1=\frac{e n_e \tau\vec{E}\cdot\vec{\mathcal{v}}}{m \mathcal{v}\left(1+i \omega \tau -i\vec{\mathcal{v}}\cdot\vec{k}\tau\right)}\frac{\partial f_{FD}}{\partial \mathcal{v}}
\end{equation}
To obtain Eq. \ref{Boltz30}, the hypothesis of isotropic unperturbed distribution $f_{FD}$ has been recalled, for which:
\begin{equation}
\vec{\nabla}_{\vec{\mathcal{v}}}f_{FD}=\frac{\vec{\mathcal{v}}}{\mathcal{v}}   \frac{\partial f_{FD}}{\partial \mathcal{v}}
\end{equation}
The general definition of current density associated with the electric carriers is given by:
\begin{equation}
    \vec{j}=-e \int f_1 \vec{\mathcal{v}} d^3 \mathcal{v}
\end{equation}
Thus, the current density component along the $z$-axis is:
\begin{equation}
    j_z= -e \int f_1 \mathcal{v}_z d^3 \mathcal{v}=  \int_0^{2\pi}d\phi\int_0^{\pi} d\theta \int_0^\infty d\mathcal{v}  \frac{e^2 n_e \tau E_z \mathcal{v} \sin{\theta}\cos{\phi} }{m \mathcal{v}\left(1+i \omega \tau -i k_z \mathcal{v} \sin{\theta}\cos{\phi} \tau-i k_r \mathcal{v}\cos{\theta}\tau\right)}\left(-\frac{\partial f_{FD}}{\partial \mathcal{v}}\right) \mathcal{v} \sin{\theta}\cos{\phi}  \mathcal{v}^2  \sin{\theta} 
\end{equation}
Therefore, it is eventually possible to demonstrate that:
\begin{equation}
    \sigma(k_r,\omega)=\frac{j_z}{E_z}= \frac{\pi n_e e^2 \tau}{m} \int_0^\infty d\mathcal{v}\int_0^\pi d\theta\sin{\theta} \frac{ \mathcal{v}^3  \sin^2{\theta}\left(-\frac{\partial f_{FD}}{\partial \mathcal{v}}\right)}{1+i  \omega\left(1- \frac{\mathcal{v}}{\beta c} \sin{\theta}\cos{\phi} \right) \tau-i k_r \mathcal{v} \cos{\theta} \tau} 
\end{equation}
By neglecting thermal effects, the derivative of the Fermi-Dirac distribution can be approximated as:
\begin{equation}
    \left(-\frac{\partial f_{FD}}{\partial \mathcal{v}}\right)\simeq \frac{3\delta(\mathcal{v}-v_F)}{4\pi v_F^3}
\end{equation}
For normal metals, $v_F<<c$, and for $\beta\simeq1$, the conductivity can be simplified as:
\begin{equation}
\label{cond2} \sigma(k_r,\omega)\simeq \frac{3 n_e e^2 \tau}{4 m  } \int_0^\pi d\theta \frac{  \sin^3{\theta}}{1+i \omega\tau-i k_r v_F \cos{\theta} \tau} 
\end{equation}
The above integral can be solved by introducing the variable
\begin{equation}
    \mathcal{s}=\frac{-i k_r v_F \tau}{1+i \omega \tau},
\end{equation}
from which the conductivity is finally obtained as:
\begin{equation}
\label{condfin}
 \sigma(k_r,\omega)=   \frac{3 n_e e^2 \tau}{4 m (1+i \omega \tau) } \left[\frac{2}{\mathcal{s}^2}+\frac{\mathcal{s}^2-1}{\mathcal{s}^3}\log{\left(\frac{1+\mathcal{s}}{1-\mathcal{s}}\right)}\right]
\end{equation}

\section{Resonance frequency}
\label{resfreqapp}
The frequency of the wakefield generated by a particle with $\beta\lesssim 1$ can be calculated by studying the poles of the function:
\begin{equation}
    \frac{ Z_0}{2\pi b  \left[\sqrt{-\frac{i\sigma_0}{\varepsilon_0\omega(1+i\omega\tau)}}+\frac{i\omega b}{ 2c}\right]}
\end{equation}
where the Bessel $I$ functions appearing in Eq. \ref{beamimped} have been approximated for large argument, consistently with the high-energy limit. However, it must be noticed that, differently from Ref. \cite{bane1996short}, no assumption is made on the value of the $\omega\tau$ parameter.
Eventually, our study of the poles leads to a resonance frequency equal to:
\begin{equation}
\begin{split}
    \omega_{res}&=Re\Bigg[\frac{i}{4\tau}-\frac{1}{4}\Bigg(-\frac{1}{\tau^2}-\frac{32\sqrt[3]{4}\sigma_0 c^2 }{\left(27 b^4 c^2\varepsilon_0^2\sigma_0+3\sqrt{81 b^8 c^4\varepsilon_0^4\sigma_0^2+3072 b^6 c^6 \varepsilon_0^3 \sigma_0^3\tau^3}\right)^{1/3}}+\\
    &+\frac{4 \left(18 b^4 c^2 \varepsilon_0^2 \sigma_0+2\sqrt{81 b^8 c^4\varepsilon_0^4\sigma_0^2+3072 b^6 c^6 \varepsilon_0^3 \sigma_0^3\tau^3}\right)^{1/3}}{3^{2/3}b^2 \varepsilon_0 \tau}\Bigg)^{1/2}+\\
    &+\frac{1}{2}\Bigg(-\frac{1}{2\tau^2}+\frac{8\sqrt[3]{4} c^2 \sigma_0}{\left(27 b^4 c^2\varepsilon_0^2\sigma_0+3\sqrt{81 b^8 c^4\varepsilon_0^4\sigma_0^2+3072 b^6 c^6 \varepsilon_0^3 \sigma_0^3\tau^3}\right)^{1/3}}+\\
    &-\frac{\left(18 b^4 c^2 \varepsilon_0^2 \sigma_0+2\sqrt{81 b^8 c^4\varepsilon_0^4\sigma_0^2+3072 b^6 c^6 \varepsilon_0^3 \sigma_0^3\tau^3}\right)^{1/3}}{3^{2/3}b^2 \varepsilon_0 \tau}+\\
    &\frac{i}{2\tau^3 \sqrt{-\frac{1}{\tau^2}-\frac{32\sqrt[3]{4}\sigma_0 c^2 }{\left(27 b^4 c^2\varepsilon_0^2\sigma_0+3\sqrt{81 b^8 c^4\varepsilon_0^4\sigma_0^2+3072 b^6 c^6 \varepsilon_0^3 \sigma_0^3\tau^3}\right)^{1/3}}+\frac{4 \left(18 b^4 c^2 \varepsilon_0^2 \sigma_0+2\sqrt{81 b^8 c^4\varepsilon_0^4\sigma_0^2+3072 b^6 c^6 \varepsilon_0^3 \sigma_0^3\tau^3}\right)^{1/3}}{3^{2/3}b^2 \varepsilon_0 \tau}}}\Bigg)^{1/2}\Bigg]
\end{split}    
\end{equation}

\section{Wave equation for an arbitrary reflection of charge carriers at the pipe's surface}
\label{reuterapp}
We retrace the Reuter and Sondheimer approach \cite{reuter1948theory} for an arbitrary value of the specular reflection probability $p$, generalizing it to the cylindrical case. First of all,
we write the Boltzmann transport equation:
\begin{equation}
\label{Boltz1}
    \frac{\partial f}{\partial t}-\frac{e}{m}\left(\Vec{E}+\vec{\mathcal{v}}\times\vec{B}\right)\cdot\vec{\nabla}_{\vec{\mathcal{v}}}f+\vec{\mathcal{v}}\cdot\vec{\nabla}_{\vec{r}}f=-\frac{f-f_0}{\tau}
\end{equation}
In the following, we'll consider non-relativistic carriers, therefore the interaction with the magnetic field can be neglected. Using the perturbative approach $f=f_0+f_1$, where $ f_0=n_0 f_{FD}$, with $n_0$ the number of carriers per unit volume in equilibrium and $f_{FD}$ the Fermi-Dirac distribution of the conductive fluid, since at the equilibrium the momenta of the quantum carriers obey such a distribution. Being $f_1$ of higher order compared to $f_0$, Eq. \ref{Boltz1} becomes:
\begin{equation}
\label{Boltz2}
    \frac{\partial f_1}{\partial t}-\frac{e}{m}\Vec{E}\cdot\vec{\nabla}_{\vec{\mathcal{v}}}f_0+\vec{\mathcal{v}}\cdot\vec{\nabla}_{\vec{r}}f_1=-\frac{f_1}{\tau}
\end{equation}
We now specialize Eq. \ref{Boltz2} to the case of a cylindrical conductive pipe, with the electric field propagating along $z$ inside the pipe, longitudinally polarized:
\begin{equation}
\label{Boltz3}
    \frac{\partial f_1}{\partial t}-\frac{e}{m}\tilde{E}_z\frac{\partial f_{0}}{\partial \mathcal{v}_z}+\mathcal{v}_r\frac{\partial f_1}{\partial r}+\mathcal{v}_z\frac{\partial f_1}{\partial z}=-\frac{f_1}{\tau}  
\end{equation}
In the frequency domain, Eq. \ref{Boltz3} can be written as:
\begin{equation}
\label{Boltz4}
    i \omega f_1-\frac{e}{m}\tilde{E}_z\frac{\partial f_{0}}{\partial \mathcal{v}_z}+\mathcal{v}_r\frac{\partial f_1}{\partial r}-\mathcal{v}_z\frac{i\omega}{\beta c}f_1=-\frac{f_1}{\tau}  
\end{equation}
Finally, one obtains:
\begin{equation}
\label{B2}
      \frac{\partial f_1}{\partial r}+ \frac{1+i \omega\tau}{\mathcal{v}_r \tau}f_1 -\frac{i\omega \mathcal{v}_z}{\beta c \mathcal{v}_r}f_1=\frac{n_0 e }{m \mathcal{v}_r }\frac{\partial f_{FD}}{\partial \mathcal{v}_z} \tilde{E}_z(r,\omega)
\end{equation}
The solution of Eq. \ref{B2} can be easily found as:
\begin{equation}
    f_1=n_0 e^{-\left(\frac{1+i \omega\tau}{\mathcal{v}_r \tau} -\frac{i\omega \mathcal{v}_z}{\beta c \mathcal{v}_r}\right)(r-b)}\left(\mathcal{F}(\vec{\mathcal{v}})+\frac{e}{m \mathcal{v}_r}\frac{\partial f_{FD}}{\partial \mathcal{v}_z}\int_{r_0}^r\tilde{E}_z(r',\omega) e^{\left(\frac{1+i \omega\tau}{\mathcal{v}_r \tau} -\frac{i\omega \mathcal{v}_z}{\beta c \mathcal{v}_r}\right)(r'-b)}\right)dr'
\end{equation}
where $\mathcal{F}(\vec{\mathcal{v}})$ is an arbitrary function of the velocity and $r_0$ is an arbitrary point, both determined by boundary conditions. 
To avoid non-physical unlimited fields, the solution for $\mathcal{v}_r<0$, i.e. for the electrons that move towards the surface, will be set as:
\begin{equation}
    f_{-}=f_0-n_0 e^{-\left(\frac{1+i \omega\tau}{\mathcal{v}_r \tau} -\frac{i\omega \mathcal{v}_z}{\beta c \mathcal{v}_r}\right)(r-b)}\frac{e}{m \mathcal{v}_r}\frac{\partial f_{FD}}{\partial \mathcal{v}_z}\int_r^{r_0\rightarrow\infty} \tilde{E}_z(r',\omega) e^{\left(\frac{1+i \omega\tau}{ \mathcal{v}_r \tau} -\frac{i\omega \mathcal{v}_z}{\beta c \mathcal{v}_r}\right)(r'-b) }dr'
\end{equation}
For the electrons that move inner inside the surface ($\mathcal{v}_r>0$),  one can assume that part of them comes from the fraction $p$ of electrons that move towards the surface ($\mathcal{v}_r<0$) and, once they reach it, they are specularly reflected $\mathcal{v}_r\rightarrow-\mathcal{v}_r$. The remaining electrons have no directionality after reflection because they are backscattered at random angles.
The boundary condition in this case, for the above reasons, is written as:
\begin{equation}
    f_{+}(r=b,r_0\rightarrow b)=p f_{-}(r=b,\mathcal{v}_r\rightarrow-\mathcal{v}_r)+(1-p)f_0
\end{equation}
\begin{equation}
\small
    f_0+n_0 \mathcal{F}(\vec{\mathcal{v}})=p f_0+p n_0 \frac{ e  }{m \mathcal{v}_r }\frac{\partial f_{FD}}{\partial \mathcal{v}_z}  \int_b^{\infty}\tilde{E}_z(r',\omega) e^{-\left(\frac{1+i \omega\tau}{ \mathcal{v}_r \tau} -\frac{i\omega \mathcal{v}_z}{\beta c \mathcal{v}_r}\right)(r'-b) }dr'+(1-p)f_0
\end{equation}
which determines $\mathcal{F}(\vec{\mathcal{v}})$ for $\mathcal{v}_r>0$:
\begin{equation}
    \mathcal{F}(\vec{\mathcal{v}})=p \frac{ e  }{m \mathcal{v}_r }\frac{\partial f_{FD}}{\partial \mathcal{v}_z}  \int_b^{\infty}\tilde{E}_z(r',\omega) e^{-\left(\frac{1+i \omega\tau}{ \mathcal{v}_r \tau} -\frac{i\omega \mathcal{v}_z}{\beta c \mathcal{v}_r}\right)(r'-b) }dr'
\end{equation}
Therefore, the dynamic solution of the Boltzmann equation for the electrons with velocity $\mathcal{v}_r>0$ is found as:
\begin{equation}
    f_{+}=f_0+ \frac{n_0  e }{m \mathcal{v}_r }\frac{\partial f_{FD}}{\partial \mathcal{v}_z}  \left[p\int_b^{\infty}\tilde{E}_z(r',\omega) e^{\left(\frac{1+i \omega\tau}{ \mathcal{v}_r \tau} -\frac{i\omega \mathcal{v}_z}{\beta c \mathcal{v}_r}\right)(r'-r)  }dr'+ \int_b^{r}\tilde{E}_z(r',\omega) e^{\left(\frac{1+i \omega\tau}{ \mathcal{v}_r \tau} -\frac{i\omega \mathcal{v}_z}{\beta c \mathcal{v}_r}\right)(r'-r) }dr'\right]
\end{equation}
After a straightforward manipulation, defining $\tilde{E}_z(r)=\tilde{E}_z(|r|)$ different from zero only for $r>0$, one can also find the expression:
\begin{equation}
    f_{+}=f_0+ \frac{n_0  e }{m \mathcal{v}_r }\frac{\partial f_{FD}}{\partial \mathcal{v}_z}  \left[p\int_0^{r}\tilde{E}_z(r',\omega) e^{\left(\frac{1+i \omega\tau}{ \mathcal{v}_r \tau} -\frac{i\omega \mathcal{v}_z}{\beta c \mathcal{v}_r}\right)(r'-r)  }dr'+ (1-p)\int_b^{r}\tilde{E}_z(r',\omega) e^{\left(\frac{1+i \omega\tau}{ \mathcal{v}_r \tau} -\frac{i\omega \mathcal{v}_z}{\beta c \mathcal{v}_r}\right)(r'-r)  }dr'\right]
\end{equation}
By its definition, we now write the electron current density as:
\begin{equation}
      j_z=-e \int \left[\delta(\mathcal{v}_r)f_{+}+\delta(-\mathcal{v}_r)f_{-}\right] \mathcal{v}_z d^3 \mathcal{v}
\end{equation}
For convenience, we can define two currents in such a way that $j_z=j_{+}+j_{-}$:
\begin{equation}
    j_{\pm}=-e \int \delta(\pm \mathcal{v}_r)f_{\pm} \mathcal{v}_z d^3 \mathcal{v}
\end{equation}
where:
\begin{equation}
\footnotesize
    j_{+}=-e \int \delta(\mathcal{v}_r) \frac{n_0  e  }{m \mathcal{v}_r }\frac{\partial f_{FD}}{\partial \mathcal{v}_z}  \left[p\int_0^{r}\tilde{E}_z(r',\omega) e^{\left(\frac{1+i \omega\tau}{ \mathcal{v}_r \tau} -\frac{i\omega \mathcal{v}_z}{\beta c \mathcal{v}_r}\right)(r'-r)  }dr'+ (1-p)\int_b^{r}\tilde{E}_z(r',\omega) e^{\left(\frac{1+i \omega\tau}{ \mathcal{v}_r \tau} -\frac{i\omega \mathcal{v}_z}{\beta c \mathcal{v}_r}\right)(r'-r)  }dr'\right] \mathcal{v}_z d^3 \mathcal{v}
\end{equation}
\begin{equation}
    j_{-}=e \int \delta(- \mathcal{v}_r) \frac{n_0  e }{m \mathcal{v}_r }\frac{\partial f_{FD}}{\partial \mathcal{v}_z} \left[ \int_r^{\infty} \tilde{E}_z(r',\omega) e^{\left(\frac{1+i \omega\tau}{ \mathcal{v}_r \tau} -\frac{i\omega \mathcal{v}_z}{\beta c \mathcal{v}_r}\right)(r-r') }dr\right] \mathcal{v}_z d^3 \mathcal{v}
\end{equation}
In the following, we work out the integrals that define the forward and backward currents. We start from $j_{-}$:
\begin{equation}
    j_{-}=-\pi e \int_0^{\infty}d\mathcal{v} \int_{\pi/2}^{\pi} d\theta  \frac{n_0  e }{m \mathcal{v} }\frac{3  \delta(\mathcal{v} -v_F)}{4\pi v_F^3} \mathcal{v} ^3 \sin^3{\theta} \sec{\theta}\left[ \int_r^{\infty} \tilde{E}_z(r',\omega) e^{\left(\frac{(1+i \omega\tau)\sec{\theta}}{ v_F \tau  } -\frac{i\omega \tan{\theta}}{\beta c }\right)(r-r') }dr\right]
\end{equation}
By integrating over $v$, it is easy to obtain:
\begin{equation}
    j_{-}=-  \int_{\pi/2}^{\pi} d\theta  \frac{3 n_0  e^2 }{4 m v_F }  \sin^3{\theta} \sec{\theta}\left[ \int_r^{\infty} \tilde{E}_z(r',\omega) e^{\left(\frac{(1+i \omega\tau)\sec{\theta}}{ v_F \tau  } -\frac{i\omega \tan{\theta}}{\beta c }\right)(r-r') }dr\right]
\end{equation}
Thus, for the opposite current, one obtains:
\begin{equation}
\begin{split}
    j_{+}&= \int_0^{\pi/2} d\theta  \frac{3 n_0  e^2 }{4 m v_F }  \sin^3{\theta} \sec{\theta}\Bigg[p\int_0^{r}\tilde{E}_z(r',\omega) e^{\left(\frac{(1+i \omega\tau)\sec{\theta}}{ v_F \tau  } -\frac{i\omega \tan{\theta}}{\beta c }\right)(r-r') }dr'+\\
    &+(1-p)\int_b^{r}\tilde{E}_z(r',\omega) e^{\left(\frac{(1+i \omega\tau)\sec{\theta}}{ v_F \tau  } -\frac{i\omega \tan{\theta}}{\beta c }\right)(r-r') }dr'\Bigg]
\end{split}    
\end{equation}
The integrals over $\theta$ appearing in the current densities can be solved analytically, giving cumbersome expressions and involving transcendent functions. \\
The wave equation for the electric field inside the metal with anomalous conductivity is, eventually:
\begin{equation}
\label{max2}
  \frac{\partial^2 \tilde{E}_z}{\partial r^2}  +     \frac{1}{r} \frac{\partial \tilde{E}_z}{\partial r}-\frac{k_z^2}{\gamma^2} \tilde{E}_z-i \mu_0 k_z \beta c (\sigma\star\tilde{E}_z)_p=0
\end{equation}
where we have defined the convolution between the electric conductivity and field as:
\begin{equation}
\label{convgenrad}
\begin{split}
(\sigma\star\tilde{E}_z)_p=j_{+}+j_{-}&=\frac{3 n_0  e^2 }{4 m v_F } \int_0^{\pi/2} d\theta\sin^3{\theta} \sec{\theta}\Bigg[p\int_0^{\infty}\tilde{E}_z(r',\omega) e^{\left(\frac{(1+i \omega\tau)\sec{\theta}}{ v_F \tau} -\frac{i\omega \tan{\theta}}{\beta c }\right)(r-r')  }dr'+\\
&+(1-p)\int_b^{\infty}\tilde{E}_z(r',\omega) e^{\left(\frac{(1+i \omega\tau)\sec{\theta}}{ v_F \tau} -\frac{i\omega \tan{\theta}}{\beta c }\right)(r-r')  }dr'\Bigg]
\end{split}
\end{equation}
Therefore, generally speaking, Eq. \ref{max2} is an integro-differential equation. In analogy to what is shown in the paper for calculating the electromagnetic field in conditions of normal conductivity, a wave equation for the electric potential may be needed. This would pave the way for finding the surface impedance and, ultimately, the coupling impedance, at least for the specular reflection of charge carriers ($p=1$). The wave equation for the electric potential is straightforwardly found as:
\begin{equation}
       \frac{\partial^2 \tilde{\phi}}{\partial r^2}+\frac{1}{r}\frac{\partial \tilde{\phi}}{\partial r}-\left(\frac{k_z^2}{\gamma^2}\tilde{\phi}+i \mu_0 k_z \beta c (\sigma\star\tilde{\phi})_p\right)=0
\end{equation}
The above equation is quite complicated due to the presence of a convolution integral, so it may not seem useful. However, we will show methods to solve it or find an expression for the surface impedance in different cases of interest. 
\subsection{Resolution of the wave equation for p=0 and surface impedance}
\label{p0dingapp}
Following the analogous procedures of Ref. \cite{reuter1948theory}, we would now like to approach the solution of the equation:
\begin{equation}
\label{eqp0}
     \frac{\partial^2 \tilde{E}_z}{\partial r^2}  +     \frac{1}{r} \frac{\partial \tilde{E}_z}{\partial r}-\frac{k_z^2}{\gamma^2} \tilde{E}_z-i \mu_0 k_z \beta c \frac{3 n_0  e^2 }{4 m v_F } \int_0^{\pi/2} d\theta\sin^3{\theta} \sec{\theta}\int_b^{\infty}\tilde{E}_z(r',\omega) e^{\left(\frac{(1+i \omega\tau)\sec{\theta}}{ v_F \tau} -\frac{i\omega \tan{\theta}}{\beta c }\right)(r-r')  }dr'=0
\end{equation}
Similarly to Ref. \cite{reuter1948theory}, it is possible to define a function $g$ such that:
\begin{equation}
    g(r,\omega)=-i \mu_0 k_z \beta c \frac{3 n_0  e^2 }{4 m v_F } \int_0^{\pi/2} d\theta\sin^3{\theta} \sec{\theta}\int_0^{b}\tilde{E}_z(r',\omega) e^{\left(\frac{(1+i \omega\tau)\sec{\theta}}{ v_F \tau} -\frac{i\omega \tan{\theta}}{\beta c }\right)(r-r')  }dr'\ for\ r<b,  \  0\ otherwise
\end{equation}
In this way, Eq. \ref{eqp0} can be recast into:
\begin{equation}
\label{eqp2}
     \frac{\partial^2 \tilde{E}_z}{\partial r^2}  +     \frac{1}{r} \frac{\partial \tilde{E}_z}{\partial r}-\frac{k_z^2}{\gamma^2} \tilde{E}_z-i \mu_0 k_z \beta c \frac{3 n_0  e^2 }{4 m v_F } \int_0^{\pi/2} d\theta\sin^3{\theta} \sec{\theta}\int_0^{\infty}\tilde{E}_z(r',\omega) e^{\left(\frac{(1+i \omega\tau)\sec{\theta}}{ v_F \tau} -\frac{i\omega \tan{\theta}}{\beta c }\right)(r-r')  }dr'=g(r,\omega)
\end{equation}
Moreover, one can define:
\begin{equation}
    F(k_r)= \int_0^{\infty} dr J_0(k_r r) r\tilde{E}_z(r,\omega)  
\end{equation}
\begin{equation}
    G(k_r)=\int_0^{ \infty} dr J_0(k_r)r g(r,\omega)
\end{equation}
\begin{equation}
\label{HHH}
    H(k_r)=\int_0^{\infty} dr J_0(k_r r)r\left[ \frac{\partial^2 \tilde{E}_z}{\partial r^2}  +     \frac{1}{r} \frac{\partial \tilde{E}_z}{\partial r}\right]= \int_0^{\infty} dr J_0(k_ r)\frac{\partial}{\partial r}\left(r \frac{\partial \tilde{E}_z}{\partial r}\right)
\end{equation}
\begin{equation}
    K(k_r)=
    \int_0^{\infty} dr J_0(k_r r)r e^{\left(\frac{(1+i \omega\tau)\sec{\theta}}{ v_F \tau} -\frac{i\omega \tan{\theta}}{\beta c }\right)r }
\end{equation}
A successive integration by parts of Eq. \ref{HHH} shows that:
\begin{equation}
\label{eqH}
    H(k_r)=F(k_r) k_r^2-\frac{\partial \tilde{E}_z}{\partial r}\Big|_{r=b^+} b J_0\left(k_r b \right)-\tilde{E}_z(b^+)  b J_1\left(k_r b \right) k_r
\end{equation}
The electric field that solves Eq. \ref{eqp0} can be found following the considerations in Ref. \cite{reuter1948theory}, based on a mathematical procedure introduced in Ref. \cite{hopf}.
However, using the simplified result shown in Ref. \cite{dingle1953anomalous} and in full analogy with it, one can obtain the following expression:
\begin{equation}
\label{eqF}
F(k_r)=\tilde{E}_z(b^+)\xi_0(k_r,\omega)=b J_1\left(k_r b \right)\frac{\tilde{E}_z(b^+)}{k_r} e^{-\frac{k_r}{\pi}\int_0^\infty\frac{\log{\left[1+\frac{\frac{\omega^2}{\gamma^2 \beta^2 c^2}+i \mu_0 \omega \sigma(\kappa,\omega)}{\kappa^2}\right]}}{\kappa^2+k_r^2}d\kappa}
\end{equation}
It is important to note that the electric field can be determined if its value at the boundary of the pipe is known. The continuity of the field at the boundary of the pipe has been discussed in the main text of this manuscript.
Inserting Eq. \ref{eqF} into \ref{eqH}, and considering the limit $k_r b>>1$, i.e. $H(k_r)\rightarrow0$, allows obtaining:
\begin{equation}
 -k_r  \tilde{E}_z(b^+) e^{-\frac{k_r}{\pi}\int_0^\infty\frac{\log{\left[1+\frac{\frac{\omega^2}{\gamma^2 \beta^2 c^2}+i \mu_0 \omega \sigma(\kappa,\omega)}{\kappa^2}\right]}}{\kappa^2+k_r^2}d\kappa} -\frac{\partial \tilde{E}_z}{\partial r}\Big|_{r=b^+}  -\tilde{E}_z(b^+)   k_r\simeq0
\end{equation}
Furthermore, the same limit allows finding an expression for the surface impedance starting from Eq. \ref{eqH}.
In fact:
\begin{equation}
\label{rrrrrr}
  \frac{\frac{\partial \tilde{E}_z}{\partial r}\Big|_{r=b^+}}{\tilde{E}_z(b^+)}\simeq   \left[k_r   e^{-\frac{k_r}{\pi}\int_0^\infty\frac{\log{\left[1+\frac{\frac{\omega^2}{\gamma^2 \beta^2 c^2}+i \mu_0 \omega \sigma(\kappa,\omega)}{\kappa^2}\right]}}{\kappa^2+k_r^2}d\kappa} -  s\right]
  \simeq-\frac{1}{\pi}\int_0^\infty\log{\left[1+\frac{\frac{\omega^2}{\gamma^2 \beta^2 c^2}+i \mu_0 \omega \sigma(k_r,\omega)}{k_r^2}\right]}dk_r
\end{equation} 
Indeed, based on Eq. \ref{eq:Z_s} and on the relation between fields and potentials, it is generally possible to write for the surface impedance:
\begin{equation}
Z_s=\frac{\tilde{E}_z(b^+,\omega)}{\tilde{H}_\varphi(b^+,\omega)}=\frac{i k_z \tilde{\phi}(b^+,\omega)-i k_z \beta c \tilde{A}_z(b^+,\omega)}{-\varepsilon_0 c^2 \frac{\partial \tilde{A}_z}{\partial r}\big|_{r=b^+}}=\frac{-i \omega Z_0  \int_0^{\infty} dk_r J_0(k_r b^+) k_r \left(1-\beta^2+\frac{i\beta^2\sigma(k_r,\omega)}{\varepsilon_0 \omega}\right) \widetilde{\phi}(k_r,\omega)}{ c \frac{\partial}{\partial r}\left(\int_0^{\infty} dk_r J_0(k_r b^+) k_r \left(\beta^2-\frac{i \beta^2\sigma(k_r,\omega)}{\varepsilon_0 \omega}\right)\widetilde{\phi}(k_r,\omega)\right)}
\end{equation}
At this point, we recall that the relation between the electric field and the potential in the reciprocal domain is:
\begin{equation}
\widetilde{E}_z(k_r,\omega)
 = \frac{i\omega }{\beta c}
\left(\frac{1}{\gamma^2}+\frac{i\beta^2\sigma(k_r,\omega)}{\varepsilon_0 \omega}\right)  \widetilde{\phi}(k_r,\omega) 
\end{equation}
For conductors good enough that $\sigma>>\varepsilon_0 \omega/\beta^2\gamma^2$,  the latter becomes approximately:
\begin{equation}
\widetilde{E}_z(k_r,\omega)
 \simeq -
\frac{\beta\sigma(k_r,\omega)}{\varepsilon_0 c} \widetilde{\phi}(k_r,\omega) 
\end{equation}
Therefore, the surface impedance can be now recast into:
\begin{equation}
\label{Zsp0approxim}
    Z_s\simeq\frac{i \omega Z_0 \int_0^{\infty} dk_r J_0(k_rb^+) k_r \left(-
\frac{\beta\sigma(k_r,\omega)}{\varepsilon_0 c}  \right) \widetilde{\phi}(k_r,\omega)}{ c \frac{\partial }{\partial r}\left(\int_0^{\infty} dk_r J_0(k_rb^+) k_r\left(-
\frac{\beta\sigma(k_r,\omega)}{\varepsilon_0 c} \right)\widetilde{\phi}(k_r,\omega)\right)}
    =\frac{i \omega Z_0  }{c}\frac{\tilde{E}_z(b^+)}{\frac{\partial \tilde{E}_z}{\partial r}\Big|_{r=b^+}}
\end{equation}
Finally, using Eqs. \ref{rrrrrr} and \ref{Zsp0approxim},  the surface impedance $Z_s$ for $p=0$ can be expressed as:
\begin{equation}
    Z_s\simeq-\frac{i \pi Z_0 \omega  }{c\int_0^\infty\log{\left[1+\frac{\frac{\omega^2}{\gamma^2 \beta^2 c^2}+i \mu_0 \omega \sigma(k_r,\omega)}{k_r^2}\right]}dk_r}
\end{equation}

\subsection{Resolution of the wave equation for p=1  and surface impedance}
\label{p1hank}
In the region occupied by the metallic pipe, i.e. for $r > b$,  the wave equation for the electric potential is written as:
\begin{equation}
\label{eqpot0}
   \frac{\partial^2 \tilde{\phi}}{\partial r^2}+\frac{1}{r}\frac{\partial \tilde{\phi}}{\partial r}-\left(\frac{k_z^2}{\gamma^2}\tilde{\phi}+i\mu_0 k_z\beta c (\sigma\star\tilde{\phi})_{p=1}\right)=0
\end{equation}
The solution of Eq. \ref{eqpot0} can be expressed through the following inverse Hankel transform:
\begin{equation}
\label{sol000}
\tilde{\phi}=\frac{q_0 C_0^\sigma(\omega)}{2\pi \varepsilon_0}\int_0^{\infty}\frac{ J_0\left(k_r r\right) k_r d k_r}{k_r^2+\bar{\eta}^2}
\end{equation}
where:
\begin{equation}
        \bar{\eta}=\frac{\omega}{\beta c}\sqrt{\frac{1}{ \gamma^2}+i\frac{\beta^2\sigma(\omega,k_r)}{\varepsilon_0\omega} }
\end{equation}
Such a result derives from the similarity with the solution in the vacuum region enclosed by the pipe. 
Indeed, it is possible to notice that if the conductivity does not depend on $k_r$, the above solution becomes:
\begin{equation}
\tilde{\phi}\simeq\frac{q_0 C_0^\sigma(\omega)}{2\pi \varepsilon_0}\int_0^{\infty}\frac{ J_0\left(k_r r\right) k_r d k_r}{k_r^2+\bar{\eta}^2}=\frac{q_0 C_0^\sigma(\omega)}{2\pi \varepsilon_0}K_0\left(\bar{\eta}r\right)
\end{equation}
which is identical to Eq. \ref{phinorm000}, as it should be. The dependence upon a $K_0$ function is the one that must be expected for the field generated by a charged particle travelling on-axis in cylindrical symmetry when the anomalous response is negligible. 
Otherwise, in the anomalous case, where the response of the medium in the radial plane is non-local, the electric potential must be calculated through the non-trivial integral in Eq. \ref{sol000}. For the electric field, by applying the Lorentz gauge in the $(k_r,\omega)$ domain, it is straightforwardly found that:
\begin{equation}
\label{efigen1}
\tilde{E}_z(r,\omega)=\frac{i \omega}{\beta c}\frac{q_0}{2\pi \varepsilon_0}C^\sigma_0(\omega)\int_0^{\infty}d k_r \left(\frac{1}{\gamma^2}+\frac{i\beta^2\sigma(k_r,\omega)}{\varepsilon_0 \omega}\right)\frac{J_0\left(k_r r\right) k_r}{ k_r^2 +\bar{\eta}^2}
\end{equation}
The magnetic field is also calculated through the electromagnetic potential:
\begin{equation}
\label{maggen1}
\tilde{H}_\varphi= -  \frac{q_0\beta c}{2\pi }C^\sigma_0(\omega)\frac{\partial}{\partial r}\int_0^{\infty}d k_r \left(1-\frac{i\sigma(k_r,\omega)}{\varepsilon_0 \omega}\right)\frac{J_0\left(k_r r\right) k_r}{ k_r^2 +\bar{\eta}^2}
\end{equation}
Both the electric and magnetic fields have been expressed through inverse Hankel transforms. The surface impedance can be calculated, for $p=1$, as:
\begin{equation}
\label{Zp1lim}
Z_{s,1}=\frac{\tilde{E}_z(b,\omega)}{\tilde{H}_\varphi(b,\omega)}=\frac{i \omega Z_0}{  c}\frac{\int_0^{\infty}d k_r \left(\frac{1}{\gamma^2}+\frac{i\beta^2\sigma(k_r,\omega)}{\varepsilon_0 \omega}\right)\frac{J_0\left(k_r b\right) k_r}{ k_r^2 +\bar{\eta}^2}}{  \int_0^{\infty}d k_r \left(\beta^2-\frac{i\beta^2\sigma(k_r,\omega)}{\varepsilon_0 \omega}\right)\frac{J_1\left(k_r b\right) k^2_r}{ k_r^2 +\bar{\eta}^2}}
\end{equation}
\subsubsection{Limit of normal conductivity for fields and surface impedance}
In the normal region, the metallic pipe response is described by the electric conductivity $\sigma(k_r,\omega)\simeq\sigma(\omega)$, therefore Eqs. \ref{efigen1} and \ref{maggen1} become:
\begin{equation}
\label{efigen2}
\tilde{E}_z(r,\omega)\simeq\frac{i \omega}{\beta c}\frac{q_0}{2\pi \varepsilon_0}C^\sigma_0(\omega)\left(\frac{1}{\gamma^2}+\frac{i\beta^2\sigma(\omega)}{\varepsilon_0 \omega}\right)\int_0^{\infty}d k_r \frac{J_0\left(k_r r\right) k_r}{ k_r^2 +\bar{\eta}^2}=\frac{i\omega }{ \beta c}
\left(\frac{1}{\gamma^2}+\frac{i\beta^2\sigma(\omega)}{\varepsilon_0 \omega}\right)\frac{q_0 }{2\pi \varepsilon_0}C_0^\sigma(\omega)K_0\left(\bar{\eta}r\right)
\end{equation}
\begin{equation}
\label{maggen2}
\tilde{H}_\varphi\simeq  -  \frac{q\beta c}{2\pi }C^\sigma_0(\omega)\left(1-\frac{i\sigma(\omega)}{\varepsilon_0 \omega}\right)\frac{\partial}{\partial r}\int_0^{\infty}d k_r \frac{J_0\left(k_r r\right) k_r}{ k_r^2 +\bar{\eta}^2}= \frac{q_0\beta c \bar{\eta}}{2\pi}C_0^\sigma(\omega)\left(1-\frac{i\sigma(\omega)}{\varepsilon_0 \omega}\right)K_1\left(\bar{\eta}r\right)
\end{equation}
Moreover, for the surface impedance of Eq. \ref{Zp1lim} in the limit $\sigma(k_r,\omega)\rightarrow\sigma(\omega)$, we obtain:
\begin{equation}
\label{limcyl0}
Z_s\simeq\frac{\left(\frac{1}{\gamma^2}+\frac{i\beta^2\sigma(\omega)}{\varepsilon_0 \omega}\right)}{\left(\beta^2-\frac{i\beta^2\sigma(\omega)}{\varepsilon_0 \omega}\right)}\frac{i \omega Z_0}{  c}\frac{\int_0^{\infty}d k_r \frac{J_0\left(k_r r\right) k_r}{ k_r^2 +\bar{\eta}^2}}{  \int_0^{\infty}d k_r \frac{J_1\left(k_r r\right) k^2_r}{ k_r^2 +\bar{\eta}^2}}=Z_0\frac{\sqrt{-\frac{1}{\beta^2\gamma^2}-i\frac{\sigma(\omega)}{\varepsilon_0\omega}}}{\left(1-\frac{i\sigma(\omega)}{\varepsilon_0 \omega}\right)}\frac{K_0\left(\bar{
 \eta}b\right)}{K_1\left(\bar{
 \eta}b\right)}
\end{equation}
which is identical to Eq. \ref{Z_s}, as to be expected.
\subsubsection{Resolution of the wave equation for p=1 (surface currents' approach) and surface impedance}
\label{rsappro}
Here, we follow the approach introduced in Ref. \cite{reuter1948theory} for the case of specular reflection of charge carriers on the metallic surface of a cylindrical beam pipe of radius $b$. Indeed, for the probability of reflection $p=1$, the wave equation for the electric field can be written as:
\begin{equation}
\label{waveq}
    \frac{\partial^2 \tilde{E}_z}{\partial r^2}  +     \frac{1}{r} \frac{\partial \tilde{E}_z}{\partial r}-\frac{k_z^2}{\gamma^2} \tilde{E}_z-i \mu_0 k_z \beta c \sigma\star \tilde{E}_z-2 \frac{\partial \tilde{E}_z}{\partial r}\Big|_{r=b^+}\delta(r=b) =0
\end{equation}
to consider the  space $r < b$ filled with another piece of the same
metal. The two pieces of metal are mirror images of one of the other. The electric field
is damped in both $r>b$ and $r<b$ directions and the boundary condition on the electric
field is:
\begin{equation}
\label{bcond}
\frac{\partial \tilde{E}_z}{\partial r}\Big|_{r=b^+}=- \frac{\partial \tilde{E}_z}{\partial r}\Big|_{r=b^-} =0
\end{equation}
Indeed, integrating Eq. \ref{waveq} upon $r$, it is possible to obtain:
\begin{equation}
\frac{\partial \tilde{E}_z}{\partial r}\Big|_{r=b^+}-\frac{\partial \tilde{E}_z}{\partial r}\Big|_{r=b^-}-2 \frac{\partial \tilde{E}_z}{\partial r}\Big|_{r=b^+} =0
\end{equation}
which is equivalent to Eq. \ref{bcond}, i.e. the boundary condition is verified.
A discontinuity of the magnetic field (the derivative of the electric field) tangent to the surface corresponds to the presence of surface electric currents.
By applying  the Hankel transform of zeroth order to Eq. \ref{waveq}, we obtain:
\begin{equation}
\label{hank1}
  -k_r^2\widetilde{E}_z-\frac{k_z^2}{\gamma^2} \widetilde{E}_z-i \mu_0 k_z \beta c \sigma(k_r,k)\widetilde{E}_z-2b\frac{\partial \tilde{E}_z}{\partial r}\Big|_{r=b^+}J_0\left(k_r b\right) =0
\end{equation}
where the Hankel transform of zeroth order of the electric field is defined as:
\begin{equation}
 \widetilde{E}_z(k_r, \omega)=\int_0^\infty dr r J_0\left(k_r r\right)  \tilde{E }_z (r,\omega)
\end{equation}
We also recall that $\omega=k_z \beta c$. By recasting Eq. \ref{hank1}, it is possible to write:
\begin{equation}
\widetilde{E}_z(k_r, \omega)=\frac{-2b\frac{\partial \tilde{E}_z}{\partial r}\Big|_{r=b^+}J_0\left(k_r b\right) }{k_r^2+\frac{\omega^2}{\beta^2\gamma^2 c^2}+i \mu_0 \omega \sigma(k_r,k)}   
\end{equation}
Finally, the inverse Hankel transform of the last equation provides the electric field inside the beam pipe (region II):
\begin{equation}
\label{fisol}
\tilde{E}_z(r,\omega)=\int_0^\infty dk_r k_r J_0\left(k_r r\right)  \widetilde{E}_z(k_r, \omega)=-2 b\frac{\partial \tilde{E}_z}{\partial r}\Big|_{r=b^+}\int_0^{\infty}\frac{k_r J_0\left(k_r b\right) J_0\left(k_r r\right) dk_r}{k_r^2+\frac{\omega^2}{\beta^2\gamma^2 c^2}+i \mu_0 \omega \sigma(k_r,k)}   
\end{equation}
In this way, while using the same approximation as in Eq. \ref{Zsp0approxim}, valid for good conductors, the surface impedance can be easily calculated as:
\begin{equation}
\label{Zp1Reut}
Z_s\simeq\frac{i \omega}{c}Z_0\frac{\tilde{E}_z(b,\omega)}{\frac{\partial  \tilde{E}_z}{\partial r}\big|_{r=b^+}}=-\frac{2 i b \omega}{c}Z_0\int_0^{\infty}\frac{k_r J^2_0\left(k_r b\right)  dk_r}{k_r^2+\frac{\omega^2}{\beta^2\gamma^2 c^2}+i \mu_0 \omega \sigma(k_r,k)} 
\end{equation}
In the limit $\sigma(k_r,\omega)\rightarrow\sigma(\omega)$, Eq. \ref{Zp1Reut} becomes:
\begin{equation}
\label{Zsp1limimag}
    Z_s\simeq -\frac{2 i \omega b  Z_0}{c} K_0\left(\bar{\eta}b\right)I_0\left(\bar{\eta}b\right)
\end{equation}
Eq. \ref{Zsp1limimag} does not immediately coincide with Eq. \ref{limcyl0}, calculated for the case $p=1$. This discrepancy shows that the approaches of Appendices \ref{p1hank} and \ref{rsappro} are not exactly equivalent. However, for large conductivities, both Eqs. \ref{limcyl0} and \ref{Zsp1limimag} tend to the classical surface impedance for (quasi-)planar surfaces:
\begin{equation}
    Z_s\simeq \frac{\omega Z_0 }{c\sqrt{\frac{\omega^2}{c^2}-\frac{\omega^2}{\beta^2 c^2}-i \frac{\mu_0 \omega \sigma_0}{1+i\omega
 \tau}}}
\end{equation}
The above result shows that, for cylindrical symmetry, the Reuter and Sondheimer approach of Ref. \cite{reuter1948theory} can be considered valid only for good conductors. Indeed,  Eq. \ref{limcyl0} is more rigorous than Eq. \ref{Zsp1limimag}. The physical reason which lies behind the above is the fact that the boundary condition in Eq. \ref{bcond}, introduced in Ref. \cite{reuter1948theory} for the planar geometry (where it is exactly valid), is only approximately valid in cylindrical geometry. Such a boundary condition can be a valid approximation in cylindrical geometry only for a strongly attenuated field inside the material, i.e. high-conductivity materials. Due to the particular topology of the hollow beam pipe, a field that does not decay rapidly near the wall of the pipe can interfere with the field located in a diametrically opposite position, making it impossible for Eq. \ref{bcond} to be exactly satisfied.

\nocite{*}

\bibliography{apssamp}

\begin{thebibliography}{30}%
\makeatletter
\providecommand \@ifxundefined [1]{%
 \@ifx{#1\undefined}
}%
\providecommand \@ifnum [1]{%
 \ifnum #1\expandafter \@firstoftwo
 \else \expandafter \@secondoftwo
 \fi
}%
\providecommand \@ifx [1]{%
 \ifx #1\expandafter \@firstoftwo
 \else \expandafter \@secondoftwo
 \fi
}%
\providecommand \natexlab [1]{#1}%
\providecommand \enquote  [1]{``#1''}%
\providecommand \bibnamefont  [1]{#1}%
\providecommand \bibfnamefont [1]{#1}%
\providecommand \citenamefont [1]{#1}%
\providecommand \href@noop [0]{\@secondoftwo}%
\providecommand \href [0]{\begingroup \@sanitize@url \@href}%
\providecommand \@href[1]{\@@startlink{#1}\@@href}%
\providecommand \@@href[1]{\endgroup#1\@@endlink}%
\providecommand \@sanitize@url [0]{\catcode `\\12\catcode `\$12\catcode `\&12\catcode `\#12\catcode `\^12\catcode `\_12\catcode `\%12\relax}%
\providecommand \@@startlink[1]{}%
\providecommand \@@endlink[0]{}%
\providecommand \url  [0]{\begingroup\@sanitize@url \@url }%
\providecommand \@url [1]{\endgroup\@href {#1}{\urlprefix }}%
\providecommand \urlprefix  [0]{URL }%
\providecommand \Eprint [0]{\href }%
\providecommand \doibase [0]{https://doi.org/}%
\providecommand \selectlanguage [0]{\@gobble}%
\providecommand \bibinfo  [0]{\@secondoftwo}%
\providecommand \bibfield  [0]{\@secondoftwo}%
\providecommand \translation [1]{[#1]}%
\providecommand \BibitemOpen [0]{}%
\providecommand \bibitemStop [0]{}%
\providecommand \bibitemNoStop [0]{.\EOS\space}%
\providecommand \EOS [0]{\spacefactor3000\relax}%
\providecommand \BibitemShut  [1]{\csname bibitem#1\endcsname}%
\let\auto@bib@innerbib\@empty
\bibitem [{\citenamefont {Vaccaro}(1966)}]{vaccaro1966longitudinal}%
  \BibitemOpen
  \bibfield  {author} {\bibinfo {author} {\bibfnamefont {V.}~\bibnamefont {Vaccaro}},\ }\href@noop {} {\emph {\bibinfo {title} {Longitudinal instability of a coasting beam above transition, due to the action of lumped discontinuities}}},\ \bibinfo {type} {Tech. Rep.}\ (\bibinfo {year} {1966})\BibitemShut {NoStop}%
\bibitem [{\citenamefont {Sessler}\ and\ \citenamefont {Vaccaro}(1967)}]{sessler1967longitudinal}%
  \BibitemOpen
  \bibfield  {author} {\bibinfo {author} {\bibfnamefont {A.~M.}\ \bibnamefont {Sessler}}\ and\ \bibinfo {author} {\bibfnamefont {V.~G.}\ \bibnamefont {Vaccaro}},\ }\href@noop {} {\emph {\bibinfo {title} {Longitudinal instabilities of azimuthally uniform beams in circular vacuum chambers with walls of arbitrary electrical properties}}},\ \bibinfo {type} {Tech. Rep.}\ (\bibinfo  {institution} {CERN},\ \bibinfo {year} {1967})\BibitemShut {NoStop}%
\bibitem [{\citenamefont {Bane}\ and\ \citenamefont {Weiland}(1984)}]{Bane_Wilson_Weiland}%
  \BibitemOpen
  \bibfield  {author} {\bibinfo {author} {\bibfnamefont {P.}~\bibnamefont {Bane}, \bibfnamefont {K.L.F.and~Wilson}}\ and\ \bibinfo {author} {\bibfnamefont {T.}~\bibnamefont {Weiland}},\ }\href@noop {} {\emph {\bibinfo {title} {Wake Fields and Wake Fields Acceleration}}},\ \bibinfo {type} {Tech. Rep.}\ (\bibinfo {year} {1984})\BibitemShut {NoStop}%
\bibitem [{\citenamefont {Palumbo}\ and\ \citenamefont {Vaccaro}(1987)}]{Palumbo_Vaccaro}%
  \BibitemOpen
  \bibfield  {author} {\bibinfo {author} {\bibfnamefont {L.}~\bibnamefont {Palumbo}}\ and\ \bibinfo {author} {\bibfnamefont {V.}~\bibnamefont {Vaccaro}},\ }\bibfield  {title} {\bibinfo {title} {Wake fields, impedances and green's function},\ }\href@noop {} {\bibfield  {journal} {\bibinfo  {journal} {CAS Advanced School on Accelerator Physics 1985}\ }\textbf {\bibinfo {volume} {CERN 87-03}} (\bibinfo {year} {1987})}\BibitemShut {NoStop}%
\bibitem [{\citenamefont {Weiland}\ and\ \citenamefont {Wanzemberg}(1991)}]{Weiland2}%
  \BibitemOpen
  \bibfield  {author} {\bibinfo {author} {\bibfnamefont {T.}~\bibnamefont {Weiland}}\ and\ \bibinfo {author} {\bibfnamefont {R.}~\bibnamefont {Wanzemberg}},\ }\href@noop {} {\emph {\bibinfo {title} {Wake Fields and Impedances}}},\ \bibinfo {type} {Tech. Rep.}\ (\bibinfo {year} {1991})\BibitemShut {NoStop}%
\bibitem [{\citenamefont {Chao}(1993)}]{Chao_1993}%
  \BibitemOpen
  \bibfield  {author} {\bibinfo {author} {\bibfnamefont {A.~W.}\ \bibnamefont {Chao}},\ }\href@noop {} {\emph {\bibinfo {title} {Physics of Collective Beam Instabilities in High Energy Accelerators}}}\ (\bibinfo  {publisher} {Wiley Series in Beam Physics and Accelerator Technology},\ \bibinfo {year} {1993})\BibitemShut {NoStop}%
\bibitem [{\citenamefont {Palumbo}\ \emph {et~al.}(1995)\citenamefont {Palumbo}, \citenamefont {Vaccaro},\ and\ \citenamefont {Zobov}}]{Palumbo_Vaccaro_Zobov}%
  \BibitemOpen
  \bibfield  {author} {\bibinfo {author} {\bibfnamefont {L.}~\bibnamefont {Palumbo}}, \bibinfo {author} {\bibfnamefont {V.}~\bibnamefont {Vaccaro}},\ and\ \bibinfo {author} {\bibfnamefont {M.}~\bibnamefont {Zobov}},\ }\bibfield  {title} {\bibinfo {title} {Wake fields and impedance},\ }\href@noop {} {\bibfield  {journal} {\bibinfo  {journal} {CAS Advanced School on Accelerator Physics [arXiv:physics/0309023]}\ }\textbf {\bibinfo {volume} {CERN 95-06}} (\bibinfo {year} {1995})}\BibitemShut {NoStop}%
\bibitem [{\citenamefont {Zotter}\ and\ \citenamefont {Keifets}(1998)}]{Zotter_Keifets}%
  \BibitemOpen
  \bibfield  {author} {\bibinfo {author} {\bibfnamefont {B.}~\bibnamefont {Zotter}}\ and\ \bibinfo {author} {\bibfnamefont {S.}~\bibnamefont {Keifets}},\ }\href@noop {} {\emph {\bibinfo {title} {Impedance and Wakes in High-Energy Particle Accelerators}}}\ (\bibinfo  {publisher} {World Scientific, Singapore},\ \bibinfo {year} {1998})\BibitemShut {NoStop}%
\bibitem [{\citenamefont {Piwinski}(1984)}]{Piwinski_flat}%
  \BibitemOpen
  \bibfield  {author} {\bibinfo {author} {\bibfnamefont {A.}~\bibnamefont {Piwinski}},\ }\href@noop {} {\emph {\bibinfo {title} {Longitudinal and transverse wake fields in flat vacuum chambers}}},\ \bibinfo {type} {Tech. Rep.}\ (\bibinfo  {institution} {DESY},\ \bibinfo {year} {1984})\BibitemShut {NoStop}%
\bibitem [{\citenamefont {Palumbo}\ and\ \citenamefont {Vaccaro}(1985)}]{Palumbo_Vaccaro_lossy}%
  \BibitemOpen
  \bibfield  {author} {\bibinfo {author} {\bibfnamefont {L.}~\bibnamefont {Palumbo}}\ and\ \bibinfo {author} {\bibfnamefont {V.}~\bibnamefont {Vaccaro}},\ }\bibfield  {title} {\bibinfo {title} {Coupling impedance between a circular beam and a lossy vacuum chamber in particle accelerators},\ }\href@noop {} {\bibfield  {journal} {\bibinfo  {journal} {Il Nuovo Cimento}\ }\textbf {\bibinfo {volume} {89}} (\bibinfo {year} {1985})}\BibitemShut {NoStop}%
\bibitem [{\citenamefont {Yokoya}(1993)}]{yokoya1993resistive}%
  \BibitemOpen
  \bibfield  {author} {\bibinfo {author} {\bibfnamefont {K.}~\bibnamefont {Yokoya}},\ }\bibfield  {title} {\bibinfo {title} {Resistive wall impedance of beam pipes of general cross section},\ }\href@noop {} {\bibfield  {journal} {\bibinfo  {journal} {Part. Accel.}\ }\textbf {\bibinfo {volume} {41}},\ \bibinfo {pages} {221} (\bibinfo {year} {1993})}\BibitemShut {NoStop}%
\bibitem [{\citenamefont {Gluckstern}\ \emph {et~al.}(1993)\citenamefont {Gluckstern}, \citenamefont {van Zeijts},\ and\ \citenamefont {Zotter}}]{gluckstern1993coupling}%
  \BibitemOpen
  \bibfield  {author} {\bibinfo {author} {\bibfnamefont {R.~L.}\ \bibnamefont {Gluckstern}}, \bibinfo {author} {\bibfnamefont {J.}~\bibnamefont {van Zeijts}},\ and\ \bibinfo {author} {\bibfnamefont {B.}~\bibnamefont {Zotter}},\ }\bibfield  {title} {\bibinfo {title} {Coupling impedance of beam pipes of general cross section},\ }\href@noop {} {\bibfield  {journal} {\bibinfo  {journal} {Physical Review E}\ }\textbf {\bibinfo {volume} {47}},\ \bibinfo {pages} {656} (\bibinfo {year} {1993})}\BibitemShut {NoStop}%
\bibitem [{\citenamefont {Piwinski}(1994)}]{Piwinski_elliptic}%
  \BibitemOpen
  \bibfield  {author} {\bibinfo {author} {\bibfnamefont {A.}~\bibnamefont {Piwinski}},\ }\href@noop {} {\emph {\bibinfo {title} {Impedances in lossy elliptical vacuum chambers}}},\ \bibinfo {type} {Tech. Rep.}\ (\bibinfo  {institution} {DESY},\ \bibinfo {year} {1994})\BibitemShut {NoStop}%
\bibitem [{\citenamefont {Ruggiero}(1996)}]{Ruggiero_1995_Arbitrary}%
  \BibitemOpen
  \bibfield  {author} {\bibinfo {author} {\bibfnamefont {F.}~\bibnamefont {Ruggiero}},\ }\bibfield  {title} {\bibinfo {title} {Resistive wall impedance as derivative of the electric capacitance for a beam pipe of arbitrary cross section},\ }\href {https://doi.org/10.1103/PhysRevE.53.2802} {\bibfield  {journal} {\bibinfo  {journal} {Phys. Rev. E}\ }\textbf {\bibinfo {volume} {53}},\ \bibinfo {pages} {2802} (\bibinfo {year} {1996})}\BibitemShut {NoStop}%
\bibitem [{\citenamefont {Bane}\ and\ \citenamefont {Sands}(1996)}]{bane1996short}%
  \BibitemOpen
  \bibfield  {author} {\bibinfo {author} {\bibfnamefont {K.~L.}\ \bibnamefont {Bane}}\ and\ \bibinfo {author} {\bibfnamefont {M.}~\bibnamefont {Sands}},\ }\bibfield  {title} {\bibinfo {title} {The short-range resistive wall wakefields},\ }in\ \href@noop {} {\emph {\bibinfo {booktitle} {AIP Conference Proceedings}}},\ Vol.\ \bibinfo {volume} {367}\ (\bibinfo {organization} {American Institute of Physics},\ \bibinfo {year} {1996})\ pp.\ \bibinfo {pages} {131--149}\BibitemShut {NoStop}%
\bibitem [{\citenamefont {Zimmermann}\ and\ \citenamefont {Oide}(2004)}]{Zimmermann_Oide}%
  \BibitemOpen
  \bibfield  {author} {\bibinfo {author} {\bibfnamefont {F.}~\bibnamefont {Zimmermann}}\ and\ \bibinfo {author} {\bibfnamefont {K.}~\bibnamefont {Oide}},\ }\bibfield  {title} {\bibinfo {title} {Resistive wall wake and impedance for non relativistic beams},\ }\href@noop {} {\bibfield  {journal} {\bibinfo  {journal} {Physical Review Special Topics, Accelerators and Beams}\ } (\bibinfo {year} {2004})}\BibitemShut {NoStop}%
\bibitem [{\citenamefont {Migliorati}\ \emph {et~al.}(2018)\citenamefont {Migliorati}, \citenamefont {Belli},\ and\ \citenamefont {Zobov}}]{migliorati_FCC}%
  \BibitemOpen
  \bibfield  {author} {\bibinfo {author} {\bibfnamefont {M.}~\bibnamefont {Migliorati}}, \bibinfo {author} {\bibfnamefont {E.}~\bibnamefont {Belli}},\ and\ \bibinfo {author} {\bibfnamefont {M.}~\bibnamefont {Zobov}},\ }\bibfield  {title} {\bibinfo {title} {Impact of the resistive wall impedance on beam dynamics in the future circular ${e}^{+}{e}^{\ensuremath{-}}$ collider},\ }\href {https://doi.org/10.1103/PhysRevAccelBeams.21.041001} {\bibfield  {journal} {\bibinfo  {journal} {Phys. Rev. Accel. Beams}\ }\textbf {\bibinfo {volume} {21}},\ \bibinfo {pages} {041001} (\bibinfo {year} {2018})}\BibitemShut {NoStop}%
\bibitem [{\citenamefont {Stupakov}(2020)}]{stupakov2020resistive}%
  \BibitemOpen
  \bibfield  {author} {\bibinfo {author} {\bibfnamefont {G.}~\bibnamefont {Stupakov}},\ }\bibfield  {title} {\bibinfo {title} {Resistive-wall wake for nonrelativistic beams revisited},\ }\href@noop {} {\bibfield  {journal} {\bibinfo  {journal} {Physical Review Accelerators and Beams}\ }\textbf {\bibinfo {volume} {23}},\ \bibinfo {pages} {094401} (\bibinfo {year} {2020})}\BibitemShut {NoStop}%
\bibitem [{\citenamefont {Stupakov}\ \emph {et~al.}(2015)\citenamefont {Stupakov}, \citenamefont {Bane}, \citenamefont {Emma},\ and\ \citenamefont {Podobedov}}]{Stupakov}%
  \BibitemOpen
  \bibfield  {author} {\bibinfo {author} {\bibfnamefont {G.}~\bibnamefont {Stupakov}}, \bibinfo {author} {\bibfnamefont {K.~L.~F.}\ \bibnamefont {Bane}}, \bibinfo {author} {\bibfnamefont {P.}~\bibnamefont {Emma}},\ and\ \bibinfo {author} {\bibfnamefont {B.}~\bibnamefont {Podobedov}},\ }\bibfield  {title} {\bibinfo {title} {Resistive wall wakefields of short bunches at cryogenic temperatures},\ }\href {https://doi.org/10.1103/PhysRevSTAB.18.034402} {\bibfield  {journal} {\bibinfo  {journal} {Phys. Rev. ST Accel. Beams}\ }\textbf {\bibinfo {volume} {18}},\ \bibinfo {pages} {034402} (\bibinfo {year} {2015})}\BibitemShut {NoStop}%
\bibitem [{\citenamefont {Wang}\ and\ \citenamefont {Qin}(2007)}]{wang2007resistive}%
  \BibitemOpen
  \bibfield  {author} {\bibinfo {author} {\bibfnamefont {N.}~\bibnamefont {Wang}}\ and\ \bibinfo {author} {\bibfnamefont {Q.}~\bibnamefont {Qin}},\ }\bibfield  {title} {\bibinfo {title} {Resistive-wall impedance of two-layer tube},\ }\href@noop {} {\bibfield  {journal} {\bibinfo  {journal} {Physical Review Special Topics—Accelerators and Beams}\ }\textbf {\bibinfo {volume} {10}},\ \bibinfo {pages} {111003} (\bibinfo {year} {2007})}\BibitemShut {NoStop}%
\bibitem [{\citenamefont {Mounet}(2012)}]{Mounet_thesis}%
  \BibitemOpen
  \bibfield  {author} {\bibinfo {author} {\bibfnamefont {N.}~\bibnamefont {Mounet}},\ }\emph {\bibinfo {title} {{The LHC Transverse Coupled-Bunch Instability}}},\ \href@noop {} {Ph.D. thesis},\ \bibinfo  {school} {Ecole Polytechnique, Lausanne} (\bibinfo {year} {2012})\BibitemShut {NoStop}%
\bibitem [{\citenamefont {Migliorati}\ \emph {et~al.}(2019)\citenamefont {Migliorati}, \citenamefont {Palumbo}, \citenamefont {Zannini}, \citenamefont {Biancacci},\ and\ \citenamefont {Vaccaro}}]{migliorati}%
  \BibitemOpen
  \bibfield  {author} {\bibinfo {author} {\bibfnamefont {M.}~\bibnamefont {Migliorati}}, \bibinfo {author} {\bibfnamefont {L.}~\bibnamefont {Palumbo}}, \bibinfo {author} {\bibfnamefont {C.}~\bibnamefont {Zannini}}, \bibinfo {author} {\bibfnamefont {N.}~\bibnamefont {Biancacci}},\ and\ \bibinfo {author} {\bibfnamefont {V.~G.}\ \bibnamefont {Vaccaro}},\ }\bibfield  {title} {\bibinfo {title} {Resistive wall impedance in elliptical multilayer vacuum chambers},\ }\href {https://doi.org/10.1103/PhysRevAccelBeams.22.121001} {\bibfield  {journal} {\bibinfo  {journal} {Phys. Rev. Accel. Beams}\ }\textbf {\bibinfo {volume} {22}},\ \bibinfo {pages} {121001} (\bibinfo {year} {2019})}\BibitemShut {NoStop}%
\bibitem [{\citenamefont {Reuter}\ and\ \citenamefont {Sondheimer}(1948)}]{reuter1948theory}%
  \BibitemOpen
  \bibfield  {author} {\bibinfo {author} {\bibfnamefont {G.}~\bibnamefont {Reuter}}\ and\ \bibinfo {author} {\bibfnamefont {E.}~\bibnamefont {Sondheimer}},\ }\bibfield  {title} {\bibinfo {title} {The theory of the anomalous skin effect in metals},\ }\href@noop {} {\bibfield  {journal} {\bibinfo  {journal} {Proceedings of the Royal Society of London. Series A. Mathematical and Physical Sciences}\ }\textbf {\bibinfo {volume} {195}},\ \bibinfo {pages} {336} (\bibinfo {year} {1948})}\BibitemShut {NoStop}%
\bibitem [{\citenamefont {Al-Khateeb}\ \emph {et~al.}(2001)\citenamefont {Al-Khateeb}, \citenamefont {Boine-Frankenheim}, \citenamefont {Hofmann},\ and\ \citenamefont {Rumolo}}]{al2001analytical}%
  \BibitemOpen
  \bibfield  {author} {\bibinfo {author} {\bibfnamefont {A.~M.}\ \bibnamefont {Al-Khateeb}}, \bibinfo {author} {\bibfnamefont {O.}~\bibnamefont {Boine-Frankenheim}}, \bibinfo {author} {\bibfnamefont {I.}~\bibnamefont {Hofmann}},\ and\ \bibinfo {author} {\bibfnamefont {G.}~\bibnamefont {Rumolo}},\ }\bibfield  {title} {\bibinfo {title} {Analytical calculation of the longitudinal space charge and resistive wall impedances in a smooth cylindrical pipe},\ }\href@noop {} {\bibfield  {journal} {\bibinfo  {journal} {Physical Review E}\ }\textbf {\bibinfo {volume} {63}},\ \bibinfo {pages} {026503} (\bibinfo {year} {2001})}\BibitemShut {NoStop}%
\bibitem [{\citenamefont {Gluckstern}(2000)}]{gluckstern2000analytic}%
  \BibitemOpen
  \bibfield  {author} {\bibinfo {author} {\bibfnamefont {R.~L.}\ \bibnamefont {Gluckstern}},\ }\bibfield  {title} {\bibinfo {title} {Analytic methods for calculating coupling impedances},\ }\href@noop {} {\bibfield  {journal} {\bibinfo  {journal} {CERN Report No. 2000-011}\ } (\bibinfo {year} {2000})}\BibitemShut {NoStop}%
\bibitem [{\citenamefont {Grosso}\ and\ \citenamefont {Parravicini}(2013)}]{grosso2013solid}%
  \BibitemOpen
  \bibfield  {author} {\bibinfo {author} {\bibfnamefont {G.}~\bibnamefont {Grosso}}\ and\ \bibinfo {author} {\bibfnamefont {G.~P.}\ \bibnamefont {Parravicini}},\ }\href@noop {} {\emph {\bibinfo {title} {Solid state physics}}}\ (\bibinfo  {publisher} {Academic press},\ \bibinfo {year} {2013})\BibitemShut {NoStop}%
\bibitem [{\citenamefont {Dingle}(1953)}]{dingle1953anomalous}%
  \BibitemOpen
  \bibfield  {author} {\bibinfo {author} {\bibfnamefont {R.}~\bibnamefont {Dingle}},\ }\bibfield  {title} {\bibinfo {title} {The anomalous skin effect and the reflectivity of metals i},\ }\href@noop {} {\bibfield  {journal} {\bibinfo  {journal} {Physica}\ }\textbf {\bibinfo {volume} {19}},\ \bibinfo {pages} {311} (\bibinfo {year} {1953})}\BibitemShut {NoStop}%
\bibitem [{\citenamefont {Podobedov}(2009)}]{podobedov2009resistive}%
  \BibitemOpen
  \bibfield  {author} {\bibinfo {author} {\bibfnamefont {B.}~\bibnamefont {Podobedov}},\ }\bibfield  {title} {\bibinfo {title} {Resistive wall wakefields in the extreme anomalous skin effect regime},\ }\href@noop {} {\bibfield  {journal} {\bibinfo  {journal} {Physical Review Special Topics—Accelerators and Beams}\ }\textbf {\bibinfo {volume} {12}},\ \bibinfo {pages} {044401} (\bibinfo {year} {2009})}\BibitemShut {NoStop}%
\bibitem [{\citenamefont {A.}(1948)}]{Leontovich}%
  \BibitemOpen
  \bibfield  {author} {\bibinfo {author} {\bibfnamefont {L.~M.}\ \bibnamefont {A.}},\ }\href@noop {} {\emph {\bibinfo {title} {Approximate boundary conditions for the electromagnetic field on the surface of a good conductor}}}\ (\bibinfo  {publisher} {House of the Academy of Sciences, Moscow},\ \bibinfo {year} {1948})\BibitemShut {NoStop}%
\bibitem [{\citenamefont {Hopf}(1934)}]{hopf}%
  \BibitemOpen
  \bibfield  {author} {\bibinfo {author} {\bibfnamefont {E.}~\bibnamefont {Hopf}},\ }\bibfield  {title} {\bibinfo {title} {Mathematical problems of radiative equilibrium},\ }\href@noop {} {\bibfield  {journal} {\bibinfo  {journal} {Cambridge Tracts in Mathematics and Mathematical Physics}\ }\textbf {\bibinfo {volume} {31}} (\bibinfo {year} {1934})}\BibitemShut {NoStop}%
\end{thebibliography}%

\end{document}